\documentclass[runningheads]{llncs}

\usepackage{eccv}

\usepackage{eccvabbrv}

\usepackage{graphicx}
\usepackage{booktabs}
\usepackage{multirow}
\usepackage{pifont}

\usepackage[accsupp]{axessibility}  

\usepackage{hyperref}

\usepackage{orcidlink}

\begin{document}

\title{SNeRV: Spectra-preserving Neural Representation for Video} 


\author{Jina Kim\thanks{Authors contribute equally.}\orcidlink{0009-0000-5988-286X} \and
Jihoo Lee$^\star$\orcidlink{0009-0005-0766-9721} \and
Je-Won Kang\thanks{Corresponding author. \\ Jihoo Lee is currently with SoC R\&D Center, LG Electronics.}\orcidlink{0000-0002-1637-9479}}

\authorrunning{J.~Kim et al.}

\institute{Department of Electronic and Electrical Engineering and Graduate Program in Smart Factory, Ewha Womans University, Seoul, South Korea 
\\
\email{qapo0106@ewhain.net, jihoo.lee@lge.com, jewonk@ewha.ac.kr}}

\maketitle

\begin{abstract}
    Neural representation for video (NeRV), which employs a neural network to parameterize video signals, introduces a novel methodology in video representations. However, existing NeRV-based methods have difficulty in capturing fine spatial details and motion patterns due to spectral bias, in which a neural network learns high-frequency (HF) components at a slower rate than low-frequency (LF) components. In this paper, we propose spectra-preserving NeRV (SNeRV) as a novel approach to enhance implicit video representations by efficiently handling various frequency components. SNeRV uses 2D discrete wavelet transform (DWT) to decompose video into LF and HF features, preserving spatial structures and directly addressing the spectral bias issue. To balance the compactness, we encode only the LF components, while HF components that include fine textures are generated by a decoder. Specialized modules, including a multi-resolution fusion unit (MFU) and a high-frequency restorer (HFR), are integrated into a backbone to facilitate the representation. Furthermore, we extend SNeRV to effectively capture temporal correlations between adjacent video frames, by casting the extension as additional frequency decomposition to a temporal domain. This approach allows us to embed spatio-temporal LF features into the network, using temporally extended up-sampling blocks (TUBs). Experimental results demonstrate that SNeRV outperforms existing NeRV models in capturing fine details and achieves enhanced reconstruction, making it a promising approach in the field of implicit video representations. The codes are available at \url{https://github.com/qwertja/SNeRV}.
\end{abstract}

\section{Introduction}
\label{sec:intro}

Implicit neural representation for video, also referred to as neural representation for video (NeRV), has attracted considerable attention as a promising method for video representations \cite{lee2022ffnerv,chen2021nerv,li2022nerv,Hnerv,zhao2023dnerv,He2023DNerV,maiya2023nirvana,kwan2024hinerv}. NeRV is used to parameterize a video signal with a neural network, in which a space-time coordinate is used as a query, and the network outputs the corresponding RGB value. As a result, a NeRV model itself can serve as a proxy for the video. This approach has proven effective in various tasks, including video compression \cite{kwan2024hinerv,lee2022ffnerv}, video interpolation \cite{jung2023anyflow, rho2022neural}, and video super-resolution \cite{chen2023motif,chen2022videoinr,lu2023learning}. 

Previous studies use pixel-wise \cite{sitzmann2020implicit, park2019deepsdf, chen2019learning} or frame-wise mapping models \cite{chen2021nerv} with various positional embedding of space-time coordinates. However, the quality of the reconstructed frame is severely degraded, when a learned decoder fails to capture the spatial details of diverse video contents. HNeRV \cite{Hnerv} utilizes a content-adaptive embedding generated by an encoder in addition to a learned decoder. Temporal correlation among adjacent video frames is exploited in the NeRV pipelines using a dynamical system \cite{zhao2023dnerv} and a decomposition of visual content and motion information \cite{He2023DNerV}. Nevertheless, capturing video representations remains challenging due to the intricate spatio-temporal dynamics in video scenes, while trying to maintain the size of a model. 

\begin{figure}[!tb]
\centering
   \includegraphics[width=\textwidth]{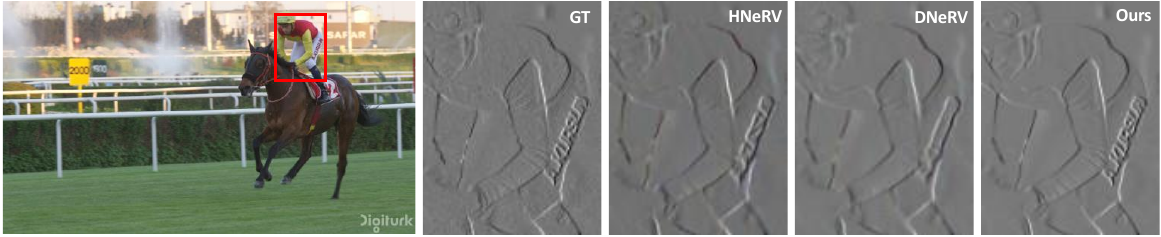}
   \hfill
\caption{Visual comparisons of the reconstructed HF coefficients of previous NeRV methods \cite{Hnerv, zhao2023dnerv} and the proposed method in ``Jockey'' sequence. Our model is designed to efficiently encode fine details, by implicitly circumventing the spectral bias problem.}
\label{fig:intro_jockey}
\end{figure}

The performance of a NeRV model is significantly affected by the characteristics of an input video \cite{zhao2023dnerv,Hnerv,li2022nerv,wang2022neural}. When a NeRV approximates an input video, the optimization is carried out to best represent the input within a budget of a model size. However, inherently, the current methods struggle to accurately capture fine spatial details and temporal patterns \cite{zhao2023dnerv,He2023DNerV,li2022nerv,Hnerv}. According to recent studies on the neural tangent kernel (NTK) \cite{ntk_spectralbias,ntk_convergence}, the learning rate of a neural network is determined by the magnitude of the eigenvalues of the target function. Because natural videos are mostly composed of low-frequency (LF) components, neural networks tend to learn the LF components more quickly, while the learning of high-frequency (HF) components proceeds at a slower rate. \cref{fig:intro_jockey} indicates that the existing NeRVs would face challenges in representing HF components due to spectral bias within the vast redundancies. 
While these behaviors are evident in the learning of implicit neural representations \cite{ntk_dictionary}, current NeRV methods have not addressed this problem effectively.

In this paper, we propose a spectra-preserving NeRV (SNeRV) that enhances the learning of implicit video representations, by efficiently processing different frequency components. First, we present a SNeRV backbone model to achieve the goal. We employ a discrete wavelet transform (DWT) to decompose a video into LF and HF features, thereby preserving their spatial structures and directly addressing the issue in the frequency domain. The model capacity is determined with the content-adaptive embeddings and the decoder sizes as \cite{Hnerv}. To represent a target in more detail, a decoder requires a larger size of a model or an embedding vector, but it compromises the model compactness. Thus, our encoder embeds only the LF component, exploiting its more suitable characteristics to learn the implicit representations. Meanwhile, the HF components including fine textures are generated by a decoder to circumvent the bias problem. This strategy enables a model to reduce redundancies and achieve a more compact representation with several specialized modules such as a multi-resolution fusion unit (MFU) and a high-frequency restorer (HFR) in a backbone model. 

On top of the backbone, we extend our model to efficiently represent temporal correlations between adjacent video frames, by casting the temporal modeling as spectral decomposition of the video frames. This is accomplished by performing an additional 1D DWT to a temporal dimension, where the motion is considered as HF components. While the backbone lacks the time analysis by maximising the capability of spectral analysis, the temporal extension considers both time and frequency domains in the analysis. By extending the time analysis, the backbone can capture temporal correlation and efficiently perform temporal modeling, well-suited for related tasks such as video interpolation. In experimental results, the backbone and its temporal extension each exhibit a performance trade-off in video regression and interpolation, while maintaining the same model capacity.

Our paper has major contributions as follows:
\begin{itemize}

\item We propose the spectra-preserving NeRV that enhances implicit video representations by efficiently processing different frequency components through specialized modules such as an MFU and a HFR.
\item We extend the backbone to capture temporal correlations between adjacent video frames, by embedding spatio-temporal LF features into the model. 
\item We evaluate the performance of the proposed methods on various video processing tasks and datasets, demonstrating their effectiveness. Our model outperforms previous approaches, emphasizing the advantages of mitigating the spectral bias problem in the INR for videos.
\end{itemize}
\section{Related Work}
\label{sec:relatedwork}

{\bf Implicit Neural Representation (INR) for Video.}
INR for video exploited multi-layer perceptron (MLP) transforming the continuous spaces of high-dimensional data into density or RGB values \cite{sitzmann2020implicit, chen2019learning}. However, the computational complexity increased substantially, as they calculated a neural function for each coordinate. To address this issue, NeRV \cite{chen2021nerv} proposed a method to use a temporal index with  positional encoding (PE) and restore up-scaled spatial information using a NeRV block with a PixelShuffle method \cite {pixelshuffle}. Based on this, E-NeRV \cite{li2022nerv} separated spatial and temporal contexts to improve the representation. PS-NeRV \cite{psnerv} further decomposed a video into a set of patches. 

Pixel-wise \cite{sitzmann2020implicit, park2019deepsdf, chen2019learning} and frame-wise \cite{chen2021nerv,psnerv} NeRV models exploited no spatial prior of a target frame, which compromised the performance. HNeRV \cite{Hnerv} proposed a hybrid model using a content-adaptive frame embedding. The embedding was created in the encoder and jointly optimized with the trained decoder parameters. By conveying the content-adaptive spatial features, it enhanced the quality of reconstructed frames. However, as the parameters were fitted to the current frame, the interpolation performance was simultaneously degraded due to the lack of temporal modeling. For this, the motion information in adjacent frames was exploited \cite{zhao2023dnerv,He2023DNerV,lee2022ffnerv,maiya2023nirvana} to achieve better results. 

\noindent{\bf Neural Networks in Frequency Domain.}
Discrete wavelet transform (DWT) decomposes a local region of an image into different frequency levels  \cite{mallat1999wavelet} and proven to be effective in various computer vision tasks \cite{huang2019wavelet,jin2020exploring,xin2020wavelet}. Multi-level space-time wavelet decomposition was employed to extract motion information using a video prediction network, while preserving spatial details \cite{jin2020exploring}. This technique was applied to image super resolution \cite{ xin2020wavelet}, enhanced with wavelet attention embeddings for video super resolution \cite{choi2021wavelet}. The DWT  typically yields sparse yet significant non-zero coefficients when representing the original signal, a property that was actively used in various directions, such as monocular depth estimation \cite{ramamonjisoa2021single} and video interpolation \cite{kong2023dynamic}.

\noindent{\bf Neural Tangent Kernel.}
Modeling the learning process of a neural network has been actively studied using kernel regression with NTK analysis \cite{ntk_spectralbias,ntk_convergence,ntk_overpar,ntk_spectralbias_2}, in which the training rate is decided by the eigenvalues of the NTK matrix. The eigenvalue spectrum of an NTK matrix explains the spectral bias of a neural network towards training speed of HF functions \cite{ntk_spectralbias_2}. As over-parameterized neural networks would be poorly suited for learning HF functions \cite{ntk_overpar}, mitigating the bias is essential for improving the performance of neural networks \cite{ntk_fast_ff,sitzmann2020implicit}. Several studies have attempted to alleviate the bias using various input mapping of a coordinate. Sinusoidal functions were modulated into an activation function \cite{sitzmann2020implicit}. Gabor-wavelets were introduced for flexible representations in space and frequency domains \cite{wire}. However, such efforts have not been explored for NeRVs to circumvent the spectral bias. Motivated by the NTK, we aim to develop compact implicit representations of diverse frequency components with a reduced number of parameters. We use the inherent sparsity of the HF wavelet transformed components in videos.

\noindent{\bf Video Compression.}
In recent years, deep learning-based video coding techniques were actively studied as alternatives to traditional codecs \cite{wiegand2003overview,sullivan2012overview,vvc}. This approach involved substituting various coding modules, such as frame restoration \cite{2020restoration}, motion prediction \cite{djelouah2019neural,lee2020convolution}, and transform and entropy coding \cite{balle}, with neural networks. DVC \cite{lu2019dvc} proposed an end-to-end video coding model. 
The NeRV models \cite{chen2021nerv,lee2022ffnerv,kwan2024hinerv,maiya2023nirvana} converted the compression tasks into model compression, using weight pruning, quantization, and entropy coding.

\begin{figure}[!tb]
\centering
   \includegraphics[width=0.9\textwidth]{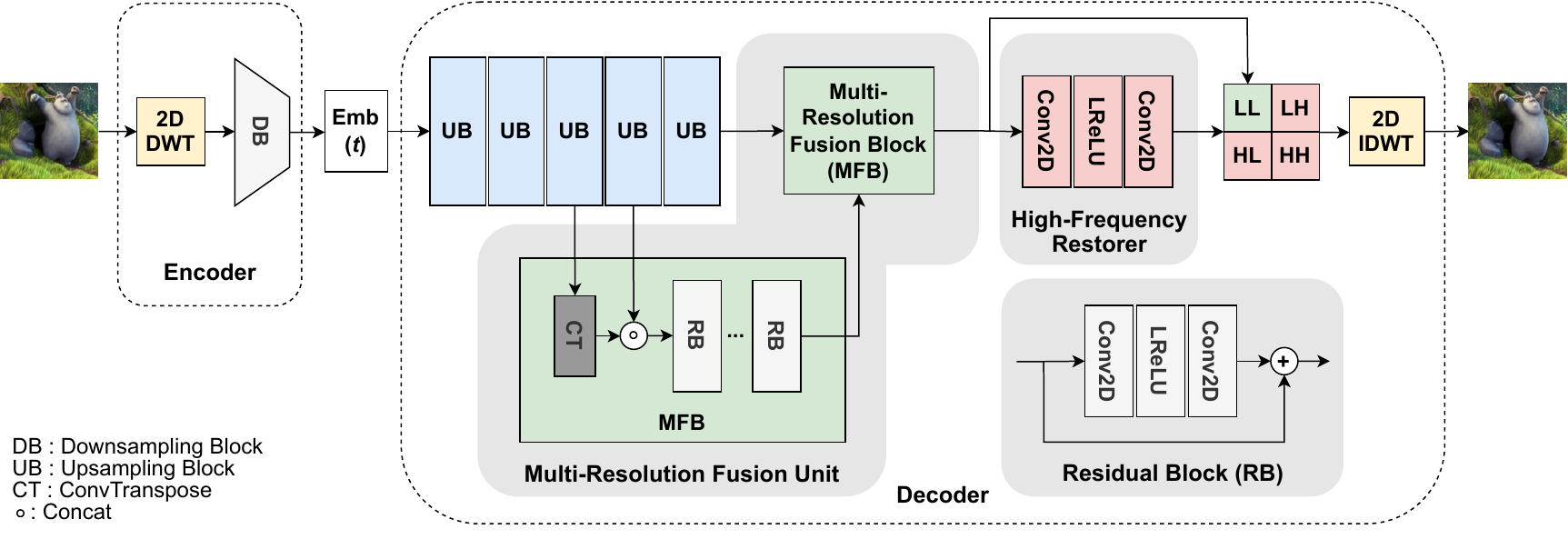}
   \hfill
\caption{SNeRV backbone encoder and decoder architectures. The encoder applies 2D DWT to extract LF and HF features and embeds only the LF feature to save parameters. The decoder uses MFU and HFR to efficiently process the LF and HF features. CT and RB refer to transposed convolution and residual blocks, respectively.}
\label{model_spat}
\end{figure}

\section{Proposed Method}
\label{sec:method}

\subsection{Overview}

Our network is designed to enhance the learning of implicit representations of video by mitigating spectral bias. We use a DWT to decompose video into LF and HF components to use an analysis-and-synthesis approach. The LF features with little variation across frames are embedded through the network, because this is inherently suitable for learning a NeRV model. The HF features are restored by a decoder. Specifically, the decoder sequentially forms the HF features with the LF ones as prior rather than directly learning the HF details. This implicit mapping enables the reallocation of the parameters to concentrate on representing LF features. Both features are optimized to consider the analysis-and-synthesis scheme, thus allowing for a more efficient representation of the HF textures and motions, based upon the stable base provided by the LF contents. 

\begin{figure}[!tb]
\centering
   \includegraphics[width=\textwidth]{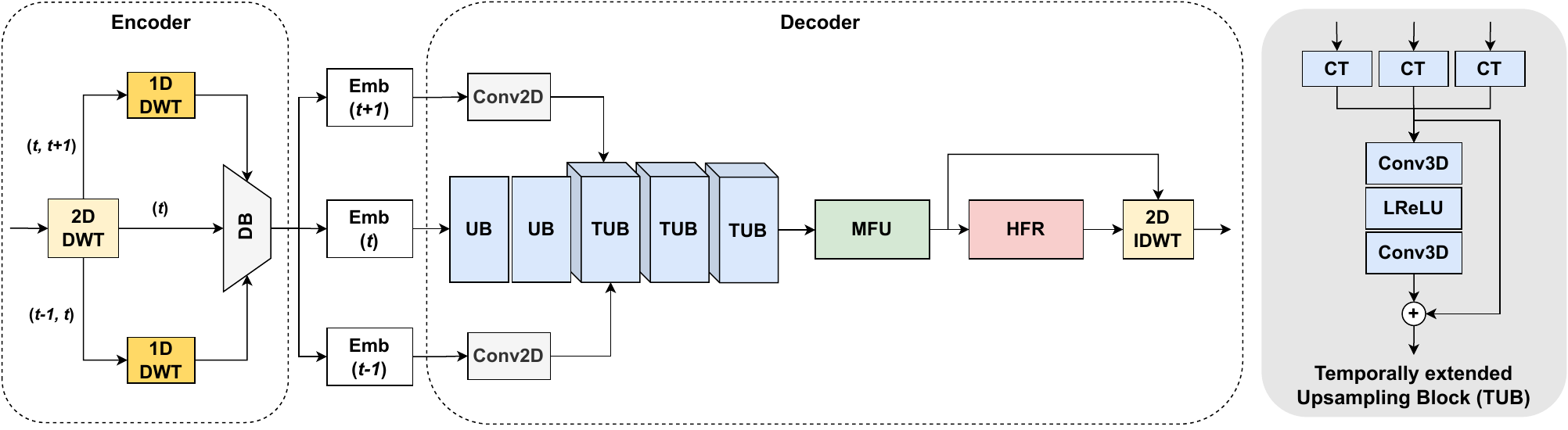}
   \hfill
\caption{Temporal extension from the backbone: the encoder uses additional 1D DWT to generate spatio-temporal embeddings. The decoder uses TUBs to address the features.}
\label{model_temp}
\end{figure}

These schemes are implemented as our backbone model of the SNeRV. 
It employs the HNeRV framework \cite{Hnerv}, in which the implicit representations are composed of a set of embedding vectors and learned decoder parameters. Based on this, we explain the novel design principles of the encoder and decoder, as shown in \cref{model_spat}. On top of that, we extend the backbone to effectively represent temporal contexts of video. We explain the extended modules to accommodate these schemes step-by-step in \cref{model_temp}. 

The differences between the backbone and its temporal extension not only clearly demonstrate their distinct architectural characteristics but also assist the performance analysis in various video processing tasks, while the backbone exhibits its superiority to the conventional methods.

\subsection{SNeRV Backbone Architecture}

{\bf Encoder Design.}
Given a video sequence $\mathbf{V}=\{I_0, I_1, \ldots, I_{T-1}\}$, in the backbone, an encoder conducts a 2D DWT along the $x-y$ spatial axes to decompose each video frame $I_t\in \mathbb{R}^{H \times W \times 3}$ into multiple wavelet sub-bands. It produces four sets of wavelet coefficients in $\mathbb{R}^{H/2 \times W/2 \times 3}$, including $C_{LL}$, $C_{LH}$, $C_{HL}$, and $C_{HH}$, at the current time-step. $C_{LL}$ contains the LF component, which often appears as a blurred and smoothed version of the original frame. The HF components, represented by $C_{LH}$, $C_{HL}$, and $C_{HH}$, encode the horizontal, vertical, and diagonal details of the image. For simplicity, we utilize the Haar WT \cite{mallat1999wavelet},  where the low pass filter is \(\phi_L = [1/\sqrt{2}, 1/\sqrt{2}]\), and the high pass filter is \(\phi_H = [1/\sqrt{2}, -1/\sqrt{2}]\). 

We input only the $C_{LL}$ into a down-sampling block (DB) to generate a content-adaptive and time-specific embedding $e_t$ whose dimension is 16$\times$2$\times$4. We use the same DB blocks as \cite{Hnerv} and keep the same dimension of $e_t$ for fair comparisons. Due to the 2D DWT, the reduced original input size alleviates the burden of parameter sizes on the decoder side. $C_{LH}$, $C_{HL}$, and $C_{HH}$ are not explicitly embedded but synthesized in the decoder.

\noindent{\bf Decoder Design.} \label{decoder}
The decoder involves the reconstruction process using $\mathbf{e}=\{e_t\}$ and learned decoder parameters. It consists of three components, including up-sampling blocks (UBs), MFU, and HFR. Firstly, UB is used to progressively increase the resolution of the embedding and reconstruct the spectral information of a target time step. We employ five NeRV blocks \cite{chen2021nerv,pixelshuffle} for the UBs. Then, the MFU fuses features from different resolutions to enrich the LF features. The HFR restores $C_{LH}$, $C_{HL}$, and $C_{HH}$ from the reconstructed $C_{LL}$. The target frame is reconstructed from an inverse DWT (IDWT), by combining all the frequency components. 
\cref{model_spat} presents the decoder architecture. We explain the MFU and HFR modules in detail as follows.

\noindent{\bf Multi-resolution Fusion Unit (MFU).}
UBs are used to sequentially reconstruct the LF features and recover the original dimension.  However, using only UBs has limits in improving the output quality of $C_{LL}$. To address this, we develop the MFU with a coarse-to-fine structure, in which the blurred outcomes are progressively refined with multi-scaled features during up-sampling. The MFU comprises two MF blocks (MFBs), each consisting of one transposed convolution and six residual blocks.

In each MFB, the current feature $m_i$ is up-sampled using the transposed convolution, and then its concatenation with the outcomes of the $i$-th UB, denoted by $u_i$, goes through the residual blocks, which is mathematically expressed as,  
\begin{gather}
    m_{i+1}=\mathcal{F}_{RB}(\mathcal{F}_{CT}(m_{i}) \circ u_{i+3}), \quad i=0,1,
\end{gather} where $\mathcal{F}_{CT}$ and $\mathcal{F}_{RB}$ are the operations of the transposed convolution and residual blocks, respectively, $\circ$ is a concatenation, and ${m_0}={u_2}$. We use $u_i$ from the last three UBs and two MFBs, empirically considering the trade-off between learning ability and the size of parameters.

\noindent{\bf High-Frequency Restorer (HFR).} Sparse representations of HF wavelet coefficients have been exploited in wavelet video compression and various computer vision researches \cite{taubman2002jpeg2000, unser2003mathematical, rho2023masked, ramamonjisoa2021single}. The HF coefficients in the flat regions of an image are nearly zero, whereas significant coefficients are found in the edges, which are essential for achieving a high quality image. Motivated by this, the HFR is designed to restore the high spatial-frequency details from the composite of LF features generated by the MFU. We employ three HFR blocks, with each comprising two convolution layers. These HFR blocks are responsible for generating $C_{LH}$, $C_{HL}$, and $C_{HH}$ from the results of $\mathcal{F}_{MFU}$. This process is shown in \cref{model_spat}. Because the synthesis process requires less parameters than the direct learning of the HF coefficients, the decoder can save the parameters and allow for more parameters to represent the LF ones.

\subsection{SNeRV Temporal Extension}
Inspired by \cite{zhao2023dnerv}, we extend the backbone to analyse the effect of considering temporal redundancies. We apply 3D (2+1D) DWT to the image sequence $\mathbf{V}$ to decompose the LF and HF components along temporal axis in addition to the existing 2D DWT. Specifically, we use two sets of adjacent frames, including $\{I_{t-1},I_t\}$ and $\{I_{t},I_{t+1}\}$ as a backward and a forward set, respectively. The 3D DWT is applied to the two image sets, in which 2D DWTs are first applied, and then 1D DWT is applied to the temporal axis. Accordingly, it produces two spatio-temporal LF components corresponding to the sets. Then, we acquire temporal embeddings $e_{\Delta t_b}$ and $e_{\Delta t_f}$ through DBs to represent the LF components of the backward and forward set, respectively.
In the decoder, $e_{\Delta t}$s are jointly up-sampled with $e_{t}$. For this, we replace the last three UBs with temporally extended UBs (TUBs), since they are used as inputs of the MFU as in the backbone. The overall architecture with temporal extension is shown in \cref{model_temp}.

\noindent{\bf Temporally Extended Up-sampling Block (TUB).} We propose the TUBs to process $e_{\Delta t_b}$ and $e_{\Delta t_f}$ in addition to $e_{t}$ using 3D convolution operations. The three inputs are up-sampled using a transposed convolution and then passed through two 3D convolution layers. The computational complexity of the 3D convolution can be significantly reduced using 2D operations. More details of our implementation can be found in the supplementary. During the consecutive 3D convolutions, the channel dimension increases by a factor of 2 and then reduces back to expand the implicit capacity of the network. In the last TUB, all three components reflecting the different time steps are merged to output the LF feature for a target at $t$. This operation allows to exploit temporal information in adjacent frames, resulting in an enhanced expressive capacity of the network over time. The outputs $C_{LL}$, $C_{LH}$, $C_{HL}$, and $C_{HH}$ for the target at $t$ are obtained by passing the results of the last TUB through the MFU and HFR, as described in \cref{decoder}, obtaining the final output by applying 2D IDWT. 

\subsection{Implementation}   The total size of the decoder parameters and embedding vectors determine the model capacity for representation. Because the performance would vary with the sizes and the extra embeddings are generated at the expense of the decoder size, we carefully explain the modified parameters to keep the total size. We set the size of $e_t$ to 16$\times$2$\times$4 and use five UBs as in  \cite{Hnerv} for fair comparisons. We use $C_{LL}$ as input that is recovered to the original dimension using IDWT. Because the IDWT replaces a UB in the last stage, we save some parameters by setting a size of a stride to the half of \cite{Hnerv, zhao2023dnerv}. This parsimony allows the parameters for MFU and HFR. In the temporal extension, we use a dimension of 3$\times$20$\times$40 for $e_{\Delta t}$. Because the size is slightly larger than that of $e_t$, in our implementation, the channel width of a decoder is reduced to maintain the overall size. We analyze the trade-offs with various temporal embedding sizes in \cref{sec:tradeoff}. More details such as stride and channel sizes can be found in the supplementary materials. 

\subsection{Loss Function}
To generate visually pleasing frames with the restored HF features, we combine both L1 and SSIM loss functions over frames, given as
\begin{equation}\label{eq:loss}
L(I_{t},\hat{I_{t}})=\frac{1}{T}\sum_{t=0}^{T-1}\alpha||I_{t}-\hat{I_{t}}||_{1} +(1-\alpha)(1-\text{SSIM}(I_{t},\hat{I_{t}})), 
\end{equation} where $\hat{I_{t}}$ is the reconstructed frame, and $I_{t}$ is the ground truth at time step $t$. SSIM ($\cdot$) calculates the SSIM score. $\alpha$ is set to 0.7 during training.  

Moreover, we compute the loss on a set of the four wavelet coefficients $\mathbf{C}$ for better frequency reconstruction. Our final loss objective is formulated as 
\begin{equation}\label{eq:loss_sum}
L_{total}=L(I_{t},\hat{I_{t}})+L(\mathbf{C}_{t},\hat{\mathbf{C}_{t}}),
\end{equation}
where $\hat{\mathbf{C}_{t}}$ and $\mathbf{C}_{t}$ are the output coefficients and the ground truth, respectively. 

\section{Experiments}
\label{sec:experiments}

\subsection{Experimental Settings}
We conduct experiments on Bunny, UVG \cite{mercat2020uvg}, and DAVIS \cite{perazzi2016benchmark} datasets. Bunny has 132 frames of 720$\times$1280 resolution. UVG is composed of 7 video sequences with 300 or 600 frames of 1080$\times$1920. We utilize 20 subsets of DAVIS datasets of 1080$\times$1920 sized frames. We used the same center-crop as in \cite{Hnerv, zhao2023dnerv}. For training, we utilized an Adam optimizer and a learning rate of $1\times 10^{-3}$ with a cosine learning rate scheduling. The batch size is set to 1. 

We use a peak signal-to-noise ratio (PSNR) and a structural similarity index measure (SSIM) for evaluation and bits-per-pixel (bpp) for compression. If none of the conditions are specified, we use 3M-sized models trained for 300 epochs in the experiments. We compare our backbone, denoted as ``Ours(B)'', and the temporal extension, denoted as ``Ours(T)'', with the state-of-the-art methods such as NeRV \cite{chen2021nerv}, E-NeRV \cite{li2022nerv}, HNeRV \cite{Hnerv}, and DNeRV \cite{zhao2023dnerv}. We train all the models, utilizing the original codes provided by the authors, and report the testing results under the same conditions. To ensure a fair comparison, we maintained the total capacity of the experimented models. To reduce the original model size from 3.8M to 3M in \cite{zhao2023dnerv}, the channel width was decreased from 92 to 70 which varies depending on the embedding sizes, resulting in a slight drop in performance compared to the original results.

\begin{table}[t]
\caption{Performance comparisons (dB) in video regression in UVG and DAVIS datasets with different sizes of resolutions.}
\centering
\resizebox{\textwidth}{!}{%
\begin{tabular}{l|l||ccccccc|c||ccccccc|c}
\toprule
\multicolumn{2}{l||}{Dataset}      & \multicolumn{8}{c||}{UVG}          & \multicolumn{8}{c}{DAVIS}            \\ \midrule
size    & method  & Beauty & Bosph  & Bee & Jockey & Ready & Shake & Yacht & Avg.  & Bike & Swan & Dance & Camel & Car   & Cows  & Twirl  & Avg.  \\ \midrule
\multirow{6}{*}{960}     & NeRV\cite{chen2021nerv}    & 33.56  & 33.28 & 37.93    & 31.31  & 26.95 & 32.67 & 28.74 & 32.06 & 25.77 & 27.22  & 25.59   & 23.82 & 28.09 & 21.75 & 25.51  & 25.39 \\
        & E-NeRV\cite{li2022nerv}  & 34.12  & 32.96 & 35.45    & 31.92  & 30.14 & 32.91 & 29.22 & 32.39 & 27.12 & 28.70  & 27.87   & 25.68 & 28.18 & 24.24 & 26.81 & 26.94 \\
        & HNeRV\cite{Hnerv}   & 34.12  & 36.38 & \underline{38.96}    & 33.58  & 29.97 & 34.15 & 31.40 & 34.08 & 30.75 & 31.77  & 28.84   & 26.59 & 33.50 & 23.85 & 28.55 & 29.12 \\
        & DNeRV\cite{zhao2023dnerv}   & 34.02  & 35.39 & 38.71    & 34.45  & 30.40 & 33.57 & 30.85 & 33.91 & 30.04 & 32.61  & 29.52   & 26.97 & 32.97 & 23.64 & 28.89 & 29.23 \\ \cmidrule{2-18}
        & Ours(B) & \textbf{34.43}  & \textbf{37.74} & 38.80    & \underline{35.69}  & \underline{33.16} & \textbf{34.52} & \textbf{33.43} & \textbf{35.40} & \textbf{33.29} & \textbf{33.83}  & \textbf{31.40}   & \underline{28.68} & \textbf{35.79} & \underline{25.14} & \textbf{30.41} & \textbf{31.22} \\
        & Ours(T) & \underline{34.14}  & \underline{37.64} & \textbf{39.11}    & \textbf{36.17}  & \textbf{33.21} & \underline{34.41} & \underline{32.71} & \underline{35.34} & \underline{31.90} & \underline{33.08}  & \underline{31.26}   & \textbf{30.14} & \underline{34.33} & \textbf{27.38} & \underline{29.93} & \underline{31.15} \\ \midrule
\multirow{6}{*}{480}  & NeRV\cite{chen2021nerv}    & 34.59  & 35.42 & 38.90    & 34.43  & 30.16 & 31.82 & 33.50 & 34.12 & 28.91 & 32.77  & 28.95   & 30.26 & 32.08 & 24.77 & 27.57 & 29.33 \\
        & E-NeRV\cite{li2022nerv}  & 35.08  & 36.63 & 39.28    & 34.62  & 31.06 & 32.43 & 33.98 & 34.73 & 29.03 & 33.35  & 29.21   & 30.98 & 34.99 & 25.76 & 26.92 & 30.04 \\
        & HNeRV\cite{Hnerv}   & 35.68  & 38.73 & \textbf{39.72}    & 37.12  & 34.38 & 34.24 & 37.63 & 36.79 & 36.07 & \underline{39.40}  & 34.62   & 35.20 & \underline{38.61} & 29.89 & 33.34  & 35.30 \\
        & DNeRV\cite{zhao2023dnerv}   & 35.30   & 38.40  & 39.50     & 37.56  & 34.18 & 33.83 & 37.14 & 36.56 & 35.66 & 39.25  & 34.22   & 35.14 & 38.35 & 29.08 & 33.24  & 34.99  \\ \cmidrule{2-18}
        & Ours(B) & \textbf{36.01}  & \underline{40.09} & 39.52    & \textbf{38.36}  & \textbf{36.90}  & \textbf{35.15} & \textbf{38.97} & \textbf{37.86} & \textbf{37.32} & \textbf{40.57}  & \textbf{36.02}   & \textbf{36.68} & \textbf{39.08} & \underline{31.78} & \textbf{35.16} & \textbf{36.66} \\
        & Ours(T) & \underline{35.78}  & \textbf{40.24} & \underline{39.61}    & \underline{38.21}  & \underline{36.77} & \underline{34.70}  & \underline{38.30} & \underline{37.66} & \underline{36.40} & 38.94  & \underline{35.59}   & \underline{36.04} & 38.17 & \textbf{32.11} & \underline{34.12} & \underline{35.91} \\ \bottomrule
\end{tabular}%
}
\label{tab:regression_uvg_davis}
\end{table}

\subsection{Performance Evaluation and Analysis}
{\bf Video Regression.}
\cref{tab:regression_uvg_davis} presents the performance of the tested methods in the tasks of video regression in UVG and DAVIS datasets with various resolutions. In the comparisons, our method presents the highest performance on average among the tested methods. In UVG dataset of 960$\times$1920 resolutions, Ours(B) improves the performance approximately by 1.32dB $\sim$ 1.46dB over HNeRV and DNeRV, respectively. In DAVIS datasets, our methods outperform the previous methods by 2.17dB $\sim$ 2.35dB. While we show the results from 7 video samples of DAVIS subsets, more results are reported in the supplementary. We observe similar results in 480$\times$960 resolutions. These results imply that our methods significantly improve the performance in video regression.

\cref{tab:regression_bunny} exhibits the performance analysis with different model sizes and training epochs in Bunny dataset. We present the results in \cref{tab:regression_size} for model sizes of 0.35M, 0.75M, 1.5M, and 3M, where the different sizes correspond to different bit budgets. The results demonstrate the superiority of our backbone model in various model capacities. Ours(T) also secures the second-best results in most scenarios. \cref{tab:regression_epoch} exhibits the performance changes, when the epochs vary from 300 to 2400 with 3M-models, implying different encoding times. Ours(B) and Ours(T) exhibit the best and second-best performance, respectively. We also show the encoding complexity of 640$\times$1280 video resolutions compared to other NeRV-based methods in \cref{tab:regression_epoch}. Despite the higher complexity involved in processing multi-scale features, Ours(B) excels in video regression, delivering superior results over other methods across all sizes and training epochs.

\begin{table}[!t]
    \caption{Performance comparisons (dB) in video regression on Bunny dataset with different model sizes and training epochs.}
    \centering
    \vspace{-2mm}
    \begin{subtable}{0.45\textwidth}
        \centering
        \caption{Performance evaluation in PSNR (dB) with different model sizes.}
        \resizebox{4.5cm}{!}{%
        {\
        \begin{tabular}{l||cccc}
        \toprule
        Model size  & 0.35M & 0.75M & 1.5M & 3.0M \\ \midrule
        NeRV\cite{chen2021nerv} & 28.16 &  29.83     & 31.88    &  33.01  \\
        E-NeRV\cite{li2022nerv} & 27.78 & 29.12  & 31.65    & 36.89  \\
        HNeRV\cite{Hnerv} & \underline{30.67} & 32.81     & 35.57    & 38.06  \\
        DNeRV\cite{zhao2023dnerv} & 29.36 & 31.97     & 35.18    & 37.55  \\ \midrule
        Ours(B)  & \textbf{30.88} & \textbf{33.25} & \textbf{36.76}    & \textbf{39.64}  \\ 
        Ours(T)  & 30.40 & \underline{33.19} & \underline{36.68} & \underline{39.09}  \\ \bottomrule
        \end{tabular}%
        }}
        \label{tab:regression_size}
    \end{subtable}
    \hfill
    \begin{subtable}{0.5\textwidth}
        \centering
        \caption{Performance evaluation with training epochs and model complexity.}
        \resizebox{5.8cm}{!}{%
        \begin{tabular}{l||ccccc||c}
        \toprule
        Epochs  & 300 & 600 & 1200 & 1800 & 2400 & MACs\\ \midrule
        NeRV\cite{chen2021nerv}  & 33.01     &  33.54    &  33.96    & 34.16    & 34.21  & 100.9G \\
        E-NeRV\cite{li2022nerv} & 36.89  & 38.69    & 39.71    & 39.90    & 39.96  & 103.3G \\
        HNeRV\cite{Hnerv} & 38.06     & 39.27    & 39.87    & 40.10    & 40.17  & 60.89G \\
        DNeRV\cite{zhao2023dnerv} & 37.55     & 38.17    & 38.55    & 38.68    & 38.74  & 48.39G \\ \midrule
        Ours(B)  & \textbf{39.64}     & \textbf{40.40}    & \textbf{40.82}    & \textbf{40.94}    & \textbf{40.99}  & 90.49G \\ 
        Ours(T)  & \underline{39.09}     & \underline{39.95}  & \underline{40.41} & \underline{40.63}    & \underline{40.68}  & 181.8G \\ \bottomrule
        \end{tabular}%
        }
        \label{tab:regression_epoch}
    \end{subtable}
    \label{tab:regression_bunny}
\end{table}

\begin{table}[!t]
\caption{Video interpolation results on UVG and DAVIS datasets.}
    \centering
    \resizebox{\textwidth}{!}{%
    \begin{tabular}{l||ccccccc|c||ccccccc|c}
    \toprule
    Dataset  & \multicolumn{8}{c||}{UVG}          & \multicolumn{8}{c}{DAVIS}            \\ \midrule
     Method & Beauty & Bosph & Bee & Jockey & Ready & Shake & Yacht & Avg. & Bike & Swan & Dance & Camel & Car & Cows & Twirl & Avg. \\ \midrule
    NeRV\cite{chen2021nerv}   & 27.74 & 29.05 & 34.50 & 18.62 & 17.54 & 28.56 & 23.54 & 25.65 & 16.70  & 19.88  & 18.03   & 17.38 & 17.55  & 17.72 & 16.28 & 17.65 \\
    E-NeRV\cite{li2022nerv}   & 26.49 & 28.31 & 35.55 & 18.07 & 17.21 & 28.22 & 23.17 & 25.29 & 16.76 & 20.02  & 18.20   & 18.25 & 16.76 & 17.25 & 16.80 & 17.72     \\
    HNeRV\cite{Hnerv}   & 30.64 & 35.57 & \underline{38.60}  & 21.08 & 21.82 & \underline{30.78} & \underline{27.31} & 29.40 & 19.26 & 20.77  & 20.33   & 19.02 & 18.51  & \underline{18.87} & 17.56 & 19.19  \\
    DNeRV\cite{zhao2023dnerv}   & \underline{30.79} & 34.07 & 38.16 & \underline{28.68} & \underline{24.76} & 30.24 & 25.97 & \underline{30.34} & \underline{20.33} & \underline{21.71}  & \underline{21.51}  & \underline{19.30}  & \underline{21.05}  & 18.11 & \underline{18.96}  & \underline{20.14}  \\ \midrule 
    Ours(B)    & 30.60 & \underline{36.44} & 38.37 & 20.81 & 21.76  & 30.70 & 27.05 & 29.37 & 19.48  & 20.80  & 20.23   & 19.10 & 18.39     & 18.82 & 17.63 & 19.21  \\
    Ours(T)   & \textbf{31.92} & \textbf{36.63} & \textbf{38.63} & \textbf{29.37} & \textbf{27.52} & \textbf{31.35} & \textbf{28.32} & \textbf{31.96} & \textbf{20.47} & \textbf{22.19}  & \textbf{21.87}   & \textbf{22.37} & \textbf{21.31}  & \textbf{21.64} & \textbf{19.12}  & \textbf{21.28}  \\ \bottomrule
    \end{tabular}%
    }
    \label{tab:interpolation}
\end{table}

\noindent{\bf Video Interpolation.}
\cref{tab:interpolation} displays the comparative studies of video interpolation in UVG and DAVIS datasets. We train the model with odd frames and test results with even frames. We observe Ours(T) and DNeRV that embed the frame differences yield better performance than the tested methods without time embeddings because the time embedding is essential to continuously represent videos over time. Ours(T) achieves the highest performance among the tested methods, which is approximately 1.14dB $\sim$ 1.62dB better than DNeRV in UVG and DAVIS datasets, respectively. 
The results indicate that Ours(T) provides efficient video representations over time.

\begin{figure}[t]
\centering
   \includegraphics[width=0.8\columnwidth]{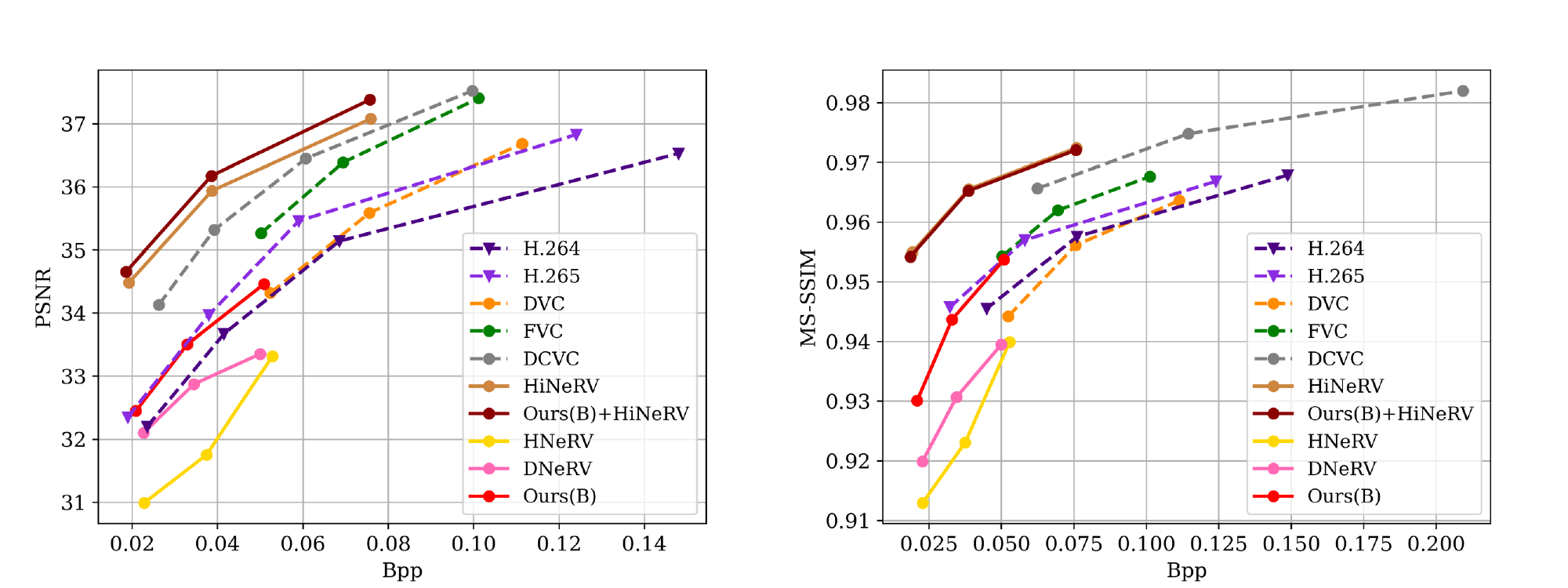}
\caption{Coding performance comparisons in UVG datasets.}
\label{fig:compression}
\end{figure}

\begin{figure}
  \begin{minipage}{\textwidth}
  \begin{minipage}[h]{0.47\textwidth}
    \captionof{table}{Video in-painting results on DAVIS dataset.}
    \centering
    \vspace{2mm}
    \resizebox{0.9\columnwidth}{!}{%
    \begin{tabular}{l||cc|c}
    \toprule
    Method & In-painting1 & In-painting2 & Avg.\\ \midrule
    HNeRV\cite{Hnerv} & 29.76        & 29.41     & 29.59 \\
    DNeRV\cite{zhao2023dnerv} & \underline{30.01}   & \underline{29.82}   & \underline{29.92}  \\ \midrule
    Ours(B)  & \textbf{30.70}   & \textbf{29.90} & \textbf{30.30}  \\ \bottomrule
    \end{tabular}%
    }
    \label{tab:Inpainting}
    \end{minipage}\vspace{-2mm}
    \hfill
  \begin{minipage}[t]{0.47\textwidth}
    \centering
    \includegraphics[width=\columnwidth]{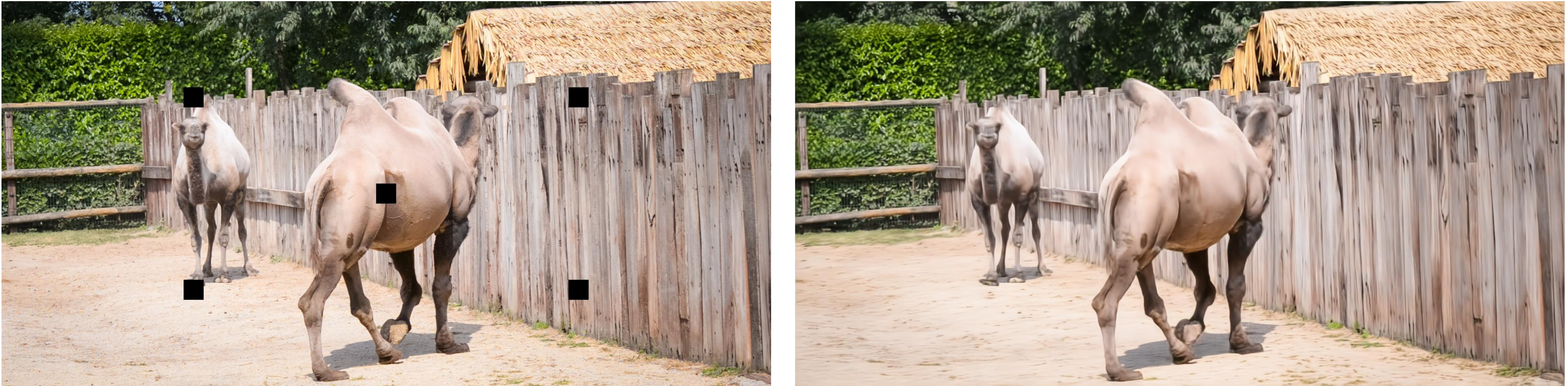}
    \captionof{figure}{Visualization of In-painting1: masked input (left) and reconstructed output (right).}
    \label{fig:vis_inpainting}
  \end{minipage}
  \end{minipage}\vspace{-2mm}
\end{figure}

\noindent{\bf Video Compression.}
We compare compression results with commercial video codecs such as H.264 \cite{wiegand2003overview} and H.265 \cite{sullivan2012overview} and learning-based codecs such as DVC \cite{lu2019dvc}, FVC \cite{hu2021fvc}, and DCVC \cite{li2021deep}. \cref{fig:compression} presents the rate-distortion curves. Our method improves coding efficiency over H.264 and all the other NeRV-based methods and achieves comparable coding gains with H.265, DVC, and FVC at low bit-rates with SSIM results. In the comparisons, we use the same coding pipeline and configurations as in \cite{zhao2023dnerv}. We adopt 10\% model pruning followed by 6-bit quantization for embeddings and 8-bit quantization for decoder parameters for all tested methods. Our methods do not apply additional techniques such as quantization-aware training (QAT) and predictive coding to improve coding efficiency for fair comparisons. {HiNeRV \cite{kwan2024hinerv} provides improved performance with specialized coding tools for NeRVs at the expense of substantial computational complexity and limited performance in video interpolation. Our method provides more fundamental changes as a backbone, which is orthogonal to the direction. Accordingly, we integrate our model into HiNeRV's coding framework and show the results denoted by Ours(B)+HiNeRV in  \cref{fig:compression}. It improves the BD-PSNR by 0.3dB compared to HiNeRV. Our supplementary results also demonstrate the generalization ability of our study.} In terms of model complexity, Ours(B) has 168G multiply–accumulates (MACs) in UVG dataset. HNeRV and DNeRV have 111.71G MACs and  69.86G MACs in the same dataset, respectively. These values are significantly smaller than those of neural network based video compression (NVC) methods such as DCVC, which has 2,268G MACs as {reported in \cite{li2021deep}}. Ours(B) has 31.3 decoding fps, which is relatively slower than 48.4 fps of HNeRV but is significantly faster than 1.75 fps of DCVC, measured in NVIDIA GeForce RTX 3090 GPU. When comparing the results, Ours(B) provides a practically reasonable trade-off between coding performance and computational complexity.

\noindent{\bf Video In-painting.}
We evaluate the performance of a video in-painting task. We use 50$\times$50 masks with five fixed points and ten random points in In-painting1 and In-painting2 \cite{zhao2023dnerv}, respectively. \cref{tab:Inpainting} presents that our model provides improved results approximately 0.38dB $\sim$ 0.71dB over the existing methods. \cref{fig:vis_inpainting} shows the visualization results of In-painting1.

\begin{figure}[!tb]
\centering
\begin{subfigure}{\textwidth}
    \includegraphics[width=\textwidth]{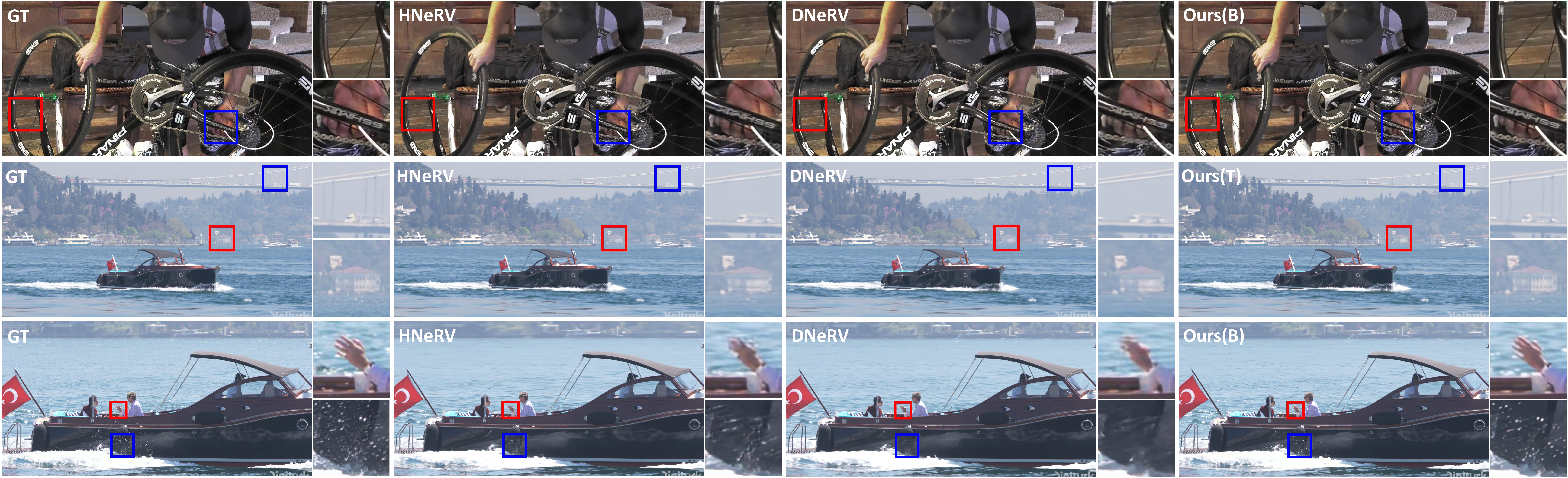}
    \caption{Visual comparisons of tested methods in the tasks of video regression. From the top, Bike, Bosph, and Yacht.}
    \label{fig:vis_regression}
\end{subfigure}
\begin{subfigure}{\textwidth}
    \includegraphics[width=\textwidth]{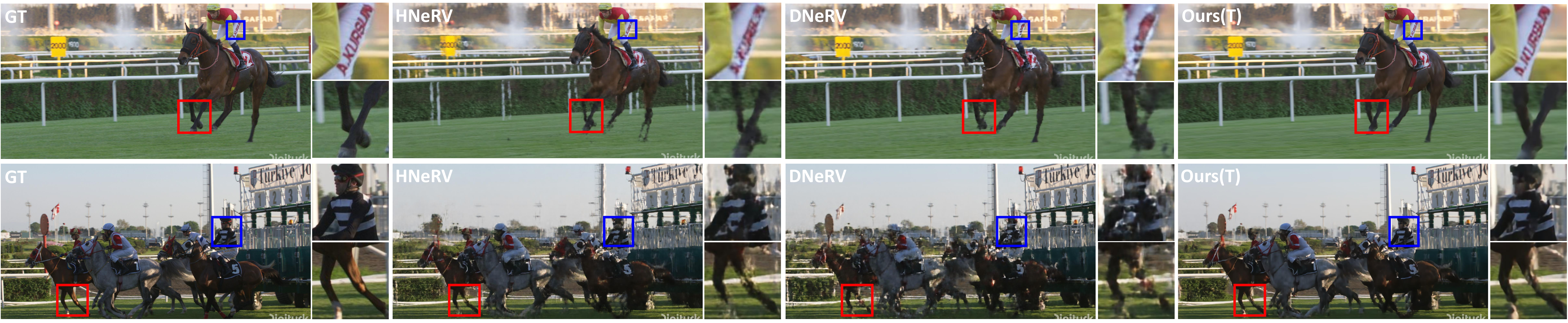}
    \caption{Visual comparisons of tested methods in the tasks of video interpolation. From the top, Jockey and Ready.}
    \label{fig:vis_interpolation}
\end{subfigure}
\caption{Visual comparisons in video regression and interpolation tasks. From the left, ground-truth (GT), HNeRV, DNeRV, and Ours.}
\label{fig:visualization1}
\end{figure}

\noindent{\bf Qualitative Performance Analysis.}
We provide the qualitative results of the tested methods in \cref{fig:visualization1}. The results illustrate that our methods provide visually pleasing results in video regression and interpolation.

\subsection{Discussion}
{\bf Trade-off between Video Regression and Time Interpolation.} 
\cref{tab:regression_uvg_davis} and \cref{tab:interpolation} display the video regression and interpolation performances, respectively. According to the results, Ours(B) outperforms Ours(T) in regression, but underperforms in interpolation, indicating a trade-off relationship. 
We explain this trade-off between time duration ($\Delta t$) and spectral band ($\Delta \omega$) in terms of learning aspects. Ours(B) sets $\Delta t$=0, lacking the time analysis but maximizing the spectral analysis, while Ours(T) compromises $\Delta t$ and $\Delta \omega$. Given the limited model parameters, learning the spatial spectral representation of a current frame could fit a reconstruction task but might compromise the generalization ability across time. In contrast, learning temporal relations can make it challenging to fit the parameters to the current time. Similar characteristics are present not only in Ours(B) and Ours(T) but also in other findings. 
DNeRV\cite{zhao2023dnerv} offers superior performance in video interpolation compared to HNeRV\cite{Hnerv}, but not in video reconstruction. This strategy in our approach signifies a new direction in analyzing and optimizing NeRV for various video processing tasks from a time and spectra perspective, which has not been conducted before.

\begin{figure}[!t]
\centering
\begin{subfigure}{0.24\columnwidth}
    \includegraphics[width=\columnwidth]{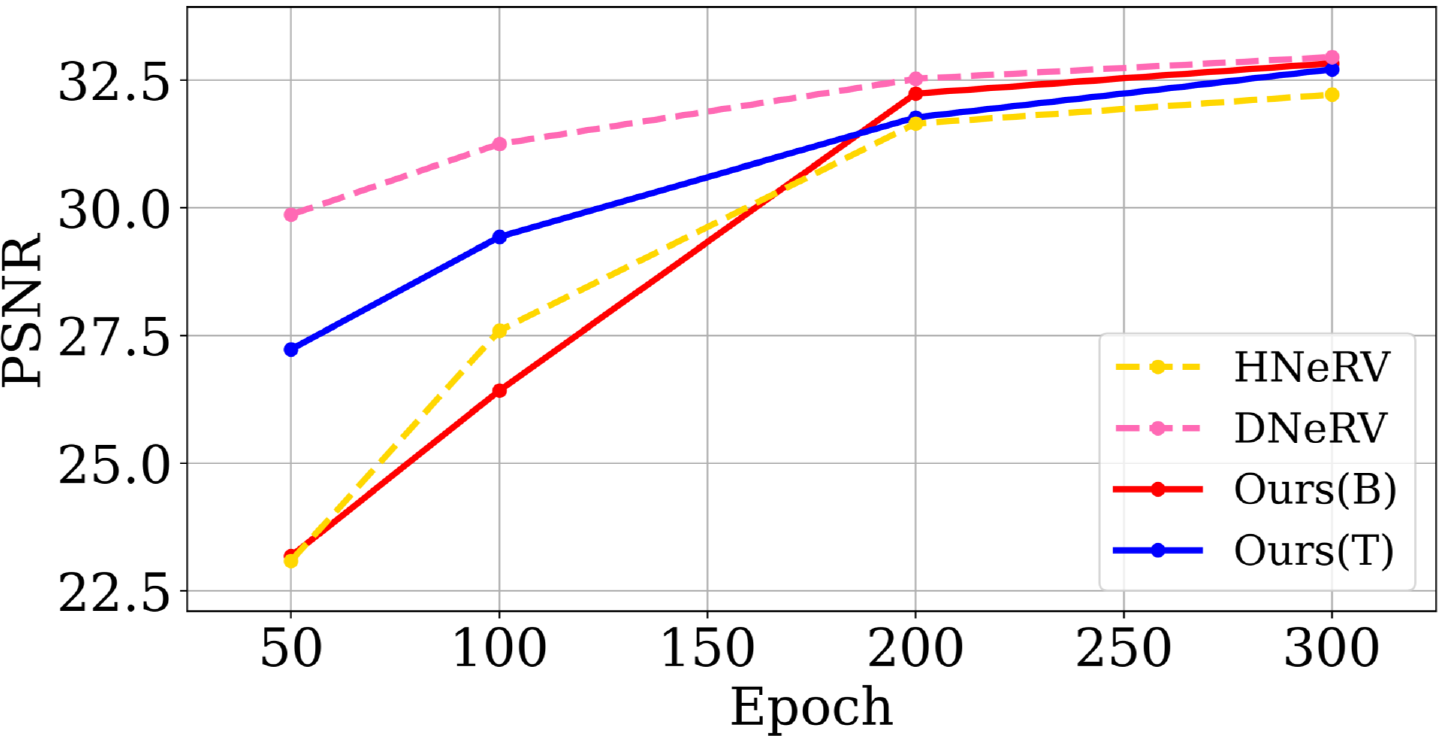}
    \caption{Low-frequency results of Jockey.}
    \label{fig:exp_freq_a}
\end{subfigure}
\begin{subfigure}{0.24\columnwidth}
    \includegraphics[width=\columnwidth]{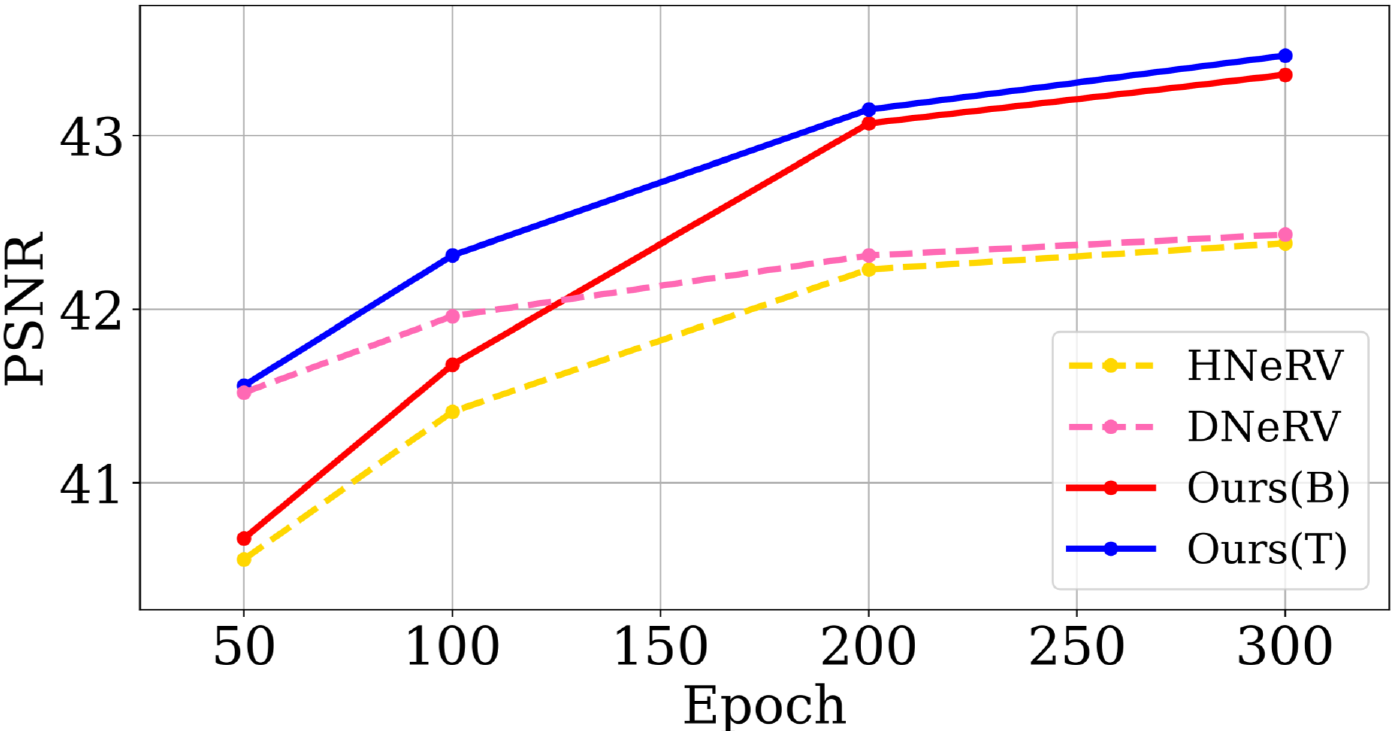}
    \caption{High-frequency results of Jockey.}
    \label{fig:exp_freq_b}
\end{subfigure}
\begin{subfigure}{0.24\columnwidth}
    \includegraphics[width=\columnwidth]{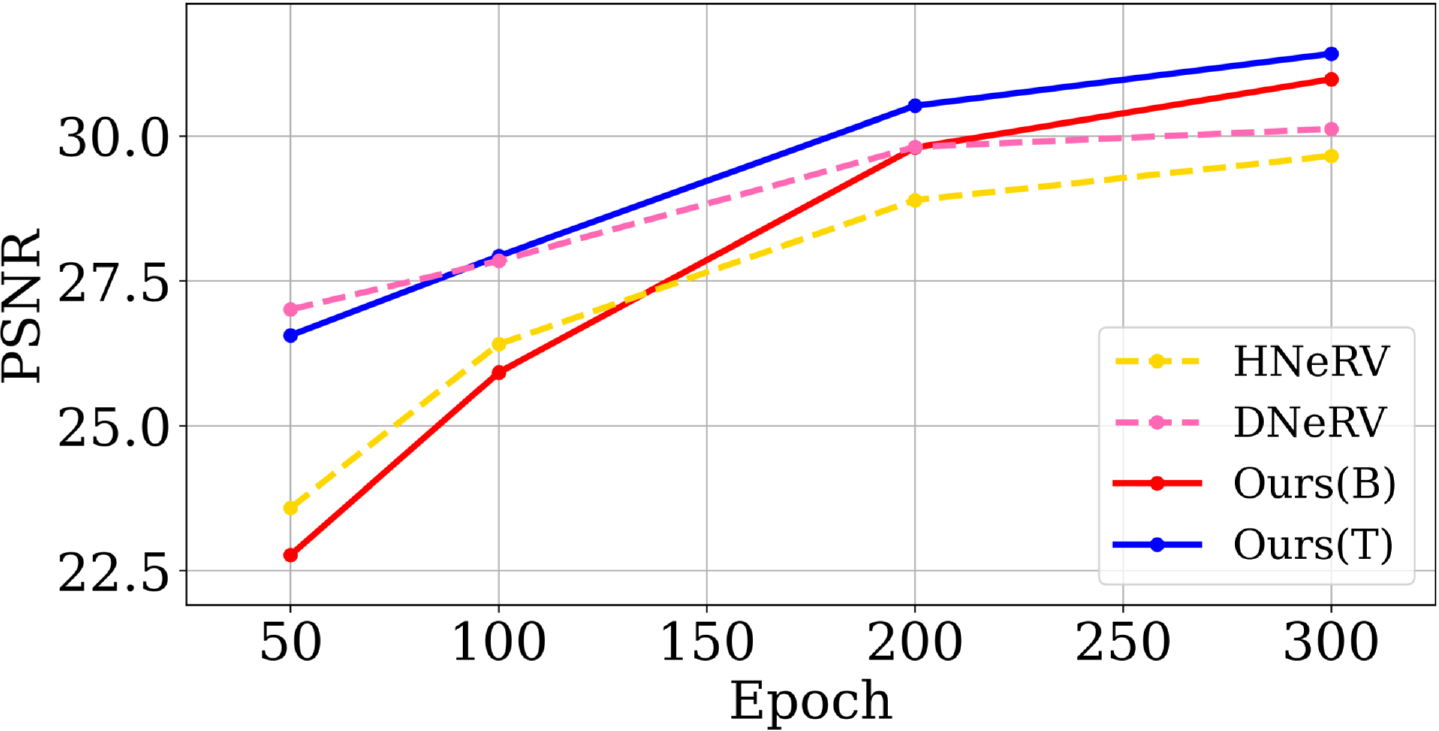}
    \caption{Low-frequency results of Ready.}
    \label{fig:exp_freq_c}
\end{subfigure}
\begin{subfigure}{0.24\columnwidth}
    \includegraphics[width=\columnwidth]{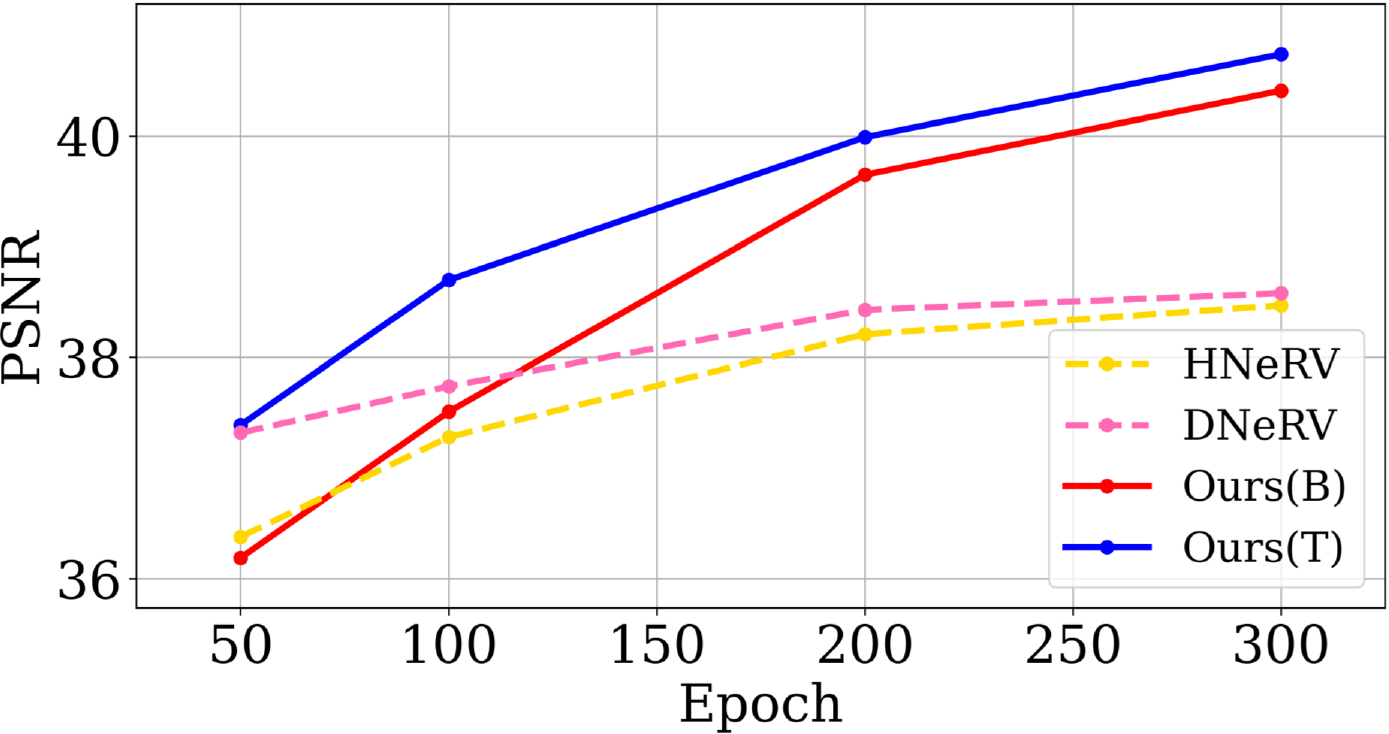}
    \caption{High-frequency results of Ready.}
    \label{fig:exp_freq_d}
\end{subfigure}
\caption{Training rates for different frequency components.}
\label{fig:freq_components}
\end{figure}

\noindent{\bf Learning Characteristics of LF and HF Components.}
\cref{fig:freq_components} displays the learning characteristics of LF and HF components separately for analysis, while they are trained simultaneously. Jockey and Ready sequences contain fine textures and motions, yielding some HF components that are challenging to represent with an implicit neural network. \cref{fig:exp_freq_a} and \cref{fig:exp_freq_c} show the results of the LF at training steps. The LFs are steadily trained to reach the convergence points producing similar PSNR values to the tested methods. Besides, Ours(T) and DNeRV exhibit better performance in the early stage, because they use larger sizes of embeddings. However, \cref{fig:exp_freq_b} and \cref{fig:exp_freq_d} show significantly different characteristics in learning HF components. While the previous methods fail to efficiently learn the HFs, our methods achieve significantly improved performance with increasing epochs. These results verify that our model could mitigate the spectral bias and successfully preserve HF components.

\subsection{Ablation Study} \label{ablation}

{\bf MFU and HFR Modules.}
\cref{tab:ablation_mfu_hfr} presents the effectiveness of the MFU and HFR. We conduct ablation tests on the Bunny dataset mostly including LF components and the Yacht sequence in UVG including comparatively larger HF components. The best performance is achieved when both the MFU and HFR are utilized. MFU achieves significantly improved performance approximately 1.20dB $\sim$ 1.88dB, demonstrating the effectiveness of the module. Using HFR, while Bunny increases the performance by 0.21dB, Yacht observes 1.62dB. This confirms that HFR is capable of restoring HF components. Further, the HFR contributes approximately 0.44dB $\sim$ 1.50dB on top of the MFU. This refinement is particularly advantageous in generating HFs that were hardly trained in previous NeRV models.

\begin{table}[!t]
    \caption{Ablation studies of the proposed modules and loss terms.}
    \centering
    \begin{subtable}{0.45\linewidth}
        \caption{MFU and HFR.}
        \centering
        \resizebox{5cm}{!}{%
        \begin{tabular}{cc||cc|cc}
        \toprule
        \multicolumn{2}{l||}{Dataset} & \multicolumn{2}{c|}{Bunny} & \multicolumn{2}{c}{Yacht} \\ \midrule
        MFU & HFR & PSNR  & SSIM & PSNR  & SSIM  \\ \midrule
        \ding{55}   & \ding{55}   & 38.00    & 0.9911 & 30.05 & 0.9323 \\
        \checkmark   & \ding{55}   & 39.20  & 0.9924 & 31.93 & 0.9458 \\
        \ding{55}   & \checkmark   & 38.21 & 0.9911 & 31.67 & 0.9516 \\
        \checkmark   & \checkmark   & 39.64 & 0.9928 & 33.43 & 0.9647 \\ \bottomrule
        \end{tabular}%
        }
        \label{tab:ablation_mfu_hfr}
        \end{subtable}
        \begin{subtable}{0.45\linewidth}
        \caption{Loss terms and TUB.}
        \centering
        \resizebox{4cm}{!}{%
        \begin{tabular}{cc|cc}
        \toprule
        \multicolumn{1}{c}{$L(I_{t},\hat{I_{t}})$} & $L(\mathbf{C}_{t},\hat{\mathbf{C}_{t}})$      & PSNR & SSIM \\ \midrule
        \multicolumn{1}{c}{} \checkmark & \ding{55} & 33.29 & 0.9600 \\
        \multicolumn{1}{c}{} \checkmark & \checkmark & 33.43 & 0.9647 \\ \bottomrule
        \toprule
        \multicolumn{2}{l|}{Temporal Block} & PSNR & SSIM \\ \midrule
        \multicolumn{2}{l|}{NeRV Block}     & 31.91 & 0.9532 \\
        \multicolumn{2}{l|}{TUB}            & 32.71 & 0.9584 \\ \bottomrule
        \end{tabular}%
        }
        \label{tab:ablation_module}
    \end{subtable}
\end{table}

\begin{figure}[!tb]
\centering
\begin{subfigure}{0.47\linewidth}
    \includegraphics[width=\columnwidth]{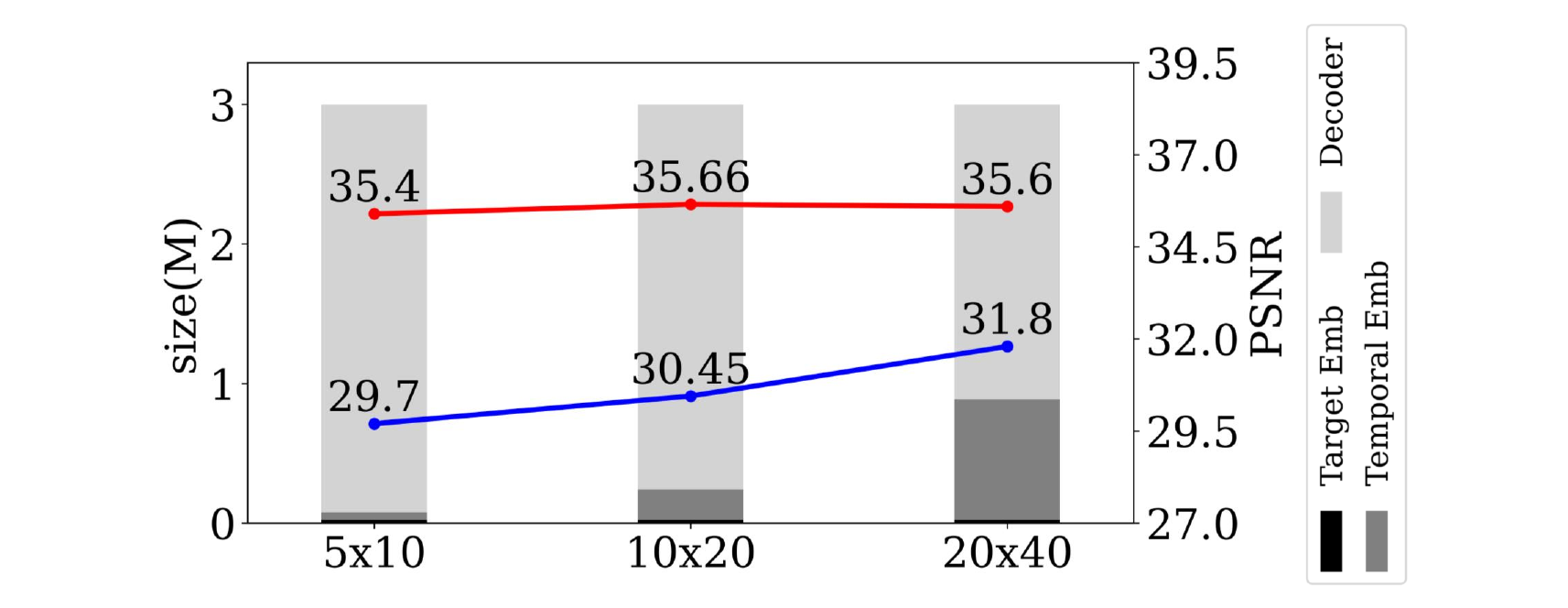}
    \caption{Varying temporal embedding size with fixed target embedding.}
    \label{fig:exp_embsize_a}
\end{subfigure}
\hfill
\begin{subfigure}{0.47\linewidth}
    \includegraphics[width=\columnwidth]{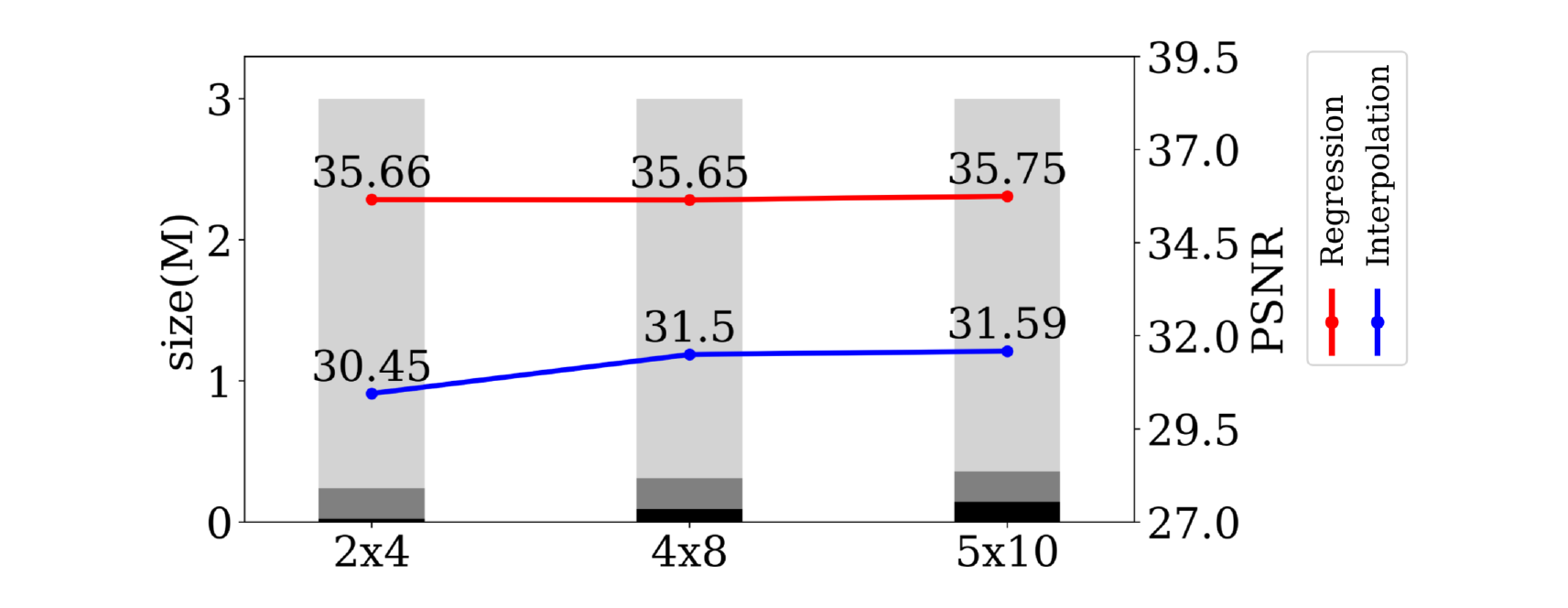}
    \caption{Varying target embedding size with fixed temporal embedding.}
    \label{fig:exp_embsize_b}
\end{subfigure}
\caption{Regression and interpolation results with different embedding sizes.}
\label{fig:exp_embsize}
\end{figure} 

\noindent{\bf Temporal Up-sampling Block.}
We evaluate the performance of the TUBs, by replacing them in Ours(T) with the conventional NeRV blocks comprising PixelShuffle and 2D convolution layers. \cref{tab:ablation_module} shows that our TUB is suitable for processing time-extended features than the 2D operations. TUB improves the PSNR around 0.80dB. 

\noindent{\bf Loss Terms.}
We conduct ablation tests on the loss function in \cref{eq:loss} using a Yacht sequence. \cref{tab:ablation_module} shows the results, when the terms are turned on or off. Both the terms contribute to the performance improvements. 

\noindent{\bf Effect of Size of Model Components.} \label{sec:tradeoff}
We analyze the performance with different sizes of target $e_{t}$ and temporal embeddings $e_{\Delta t}$s, with the size of the decoder varying according to these sizes. \cref{fig:exp_embsize} presents the performance variations, exploring the impacts of the ratios. The regression performance is reliable in all the cases, when the ratios are different. In interpolation, the size of the temporal embedding greatly affects the performance, because it hints the temporal relations among video frames.

\section{Conclusion}
\label{sec:conclusion}

In this paper, we proposed the spectra-preserving neural representation for video as an efficient baseline framework. SNeRV successfully learned LF components, while reconstructing fine HF details within a limited capacity of parameters, and surpassed the existing NeRV-based methods in various video-related tasks. More compact and efficient architecture to process the high-frequency components considering time-frequency trade-off will be done as a future work.

\section*{Acknowledgments}
This work was partly supported by Institute of Information $\&$ communications Technology Planning $\&$ Evaluation (IITP) grant funded by the Korea government(MSIT) (No.RS-2021-II212068 and RS-2022-00167169) and was partly supported by the NRF grant funded by MSIT (No.NRF-2022R1A2C4002052).

%
%
\bibliographystyle{splncs04}
\bibliography{main}

\renewcommand{\thesection}{\Alph{section}}
\clearpage
\setcounter{page}{1}
\setcounter{section}{0}
\title{Supplementary Material for SNeRV: Spectra-preserving Neural Representation for Video}

\author{Jina Kim$^\star$\inst{1}\orcidlink{0000-1111-2222-3333} \and
Jihoo Lee$^\star$\inst{1,2}\orcidlink{1111-2222-3333-4444} \and
Je-Won Kang$^\star$$^\star$\inst{1}\orcidlink{2222--3333-4444-5555}}

\authorrunning{J.~Kim et al.}
\titlerunning{SNeRV: Spectra-preserving Neural Representation for Video}

\institute{Dept. of Electronic and Electrical Engineering and Graduate Program in Smart Factory, Ewha W. University, Seoul, South Korea \and
SoC R\&D Center, LG Electronics \\
\email{qapo0106@ewhain.net, jihoo.lee@lge.com, jewonk@ewha.ac.kr}
}
\maketitle

\section{Architecture Details}
\subsection{Architecture Details of SNeRV}
The stride and the channel size for decoders differ due to the constraint of the total model size, when the resolutions of input video sequences are different. For the decoder, the reduction rate $r=1.2$ is used for each UBs as \cite{Hnerv}. In detail, the input channel size for each UB is reduced by the factor of reduction rate $r$. This makes the input channel of initial UB, which is denoted as $C_0$, to determine the overall channel sizes of the decoder. The SNeRV decoder architecture details are provided in \cref{tab:SNeRV_details}, in which the specific parameters vary with the input resolutions to keep the constraint of a total size of 3M. $N_{RB}$ denotes the number of RBs in a single MFB. Additionally, we use 0.1 negative slope for a LeakyReLU activation function.

\begin{table}[!ht]
\centering
    \caption{SNeRV decoder architecture details.}
    \resizebox{5cm}{!}{%
    \begin{tabular}{l||ccc}
    \toprule
    Input size & Strides     & $C_{0}$  & $N_{RB}$ \\ \midrule
    640$\times$1280   & 5, 4, 2, 2, 2 & 111 & 6                 \\
    480$\times$960    & 5, 3, 2, 2, 2 & 119 & 6                 \\
    960$\times$1920   & 5, 4, 3, 2, 2 & 100 & 6                 \\  \bottomrule
    \end{tabular}
    }
    \label{tab:SNeRV_details}
\end{table}

\subsection{Architecture Details of Temporally Extended SNeRV}
We present the detailed decoder architecture of temporally extended SNeRV in \cref{tab:SNeRV_Tdec_details}. We make the total size as same as the SNeRV. Also, for the first two UBs in the temporal extension, we use two consecutive 2D $3\times3$ convolutions with a residual connection instead of NeRV blocks. Additionally, we report the stride of DBs in the encoder, which generates different sizes of temporal embeddings in  \cref{tab:SNeRV_Tenc_details}. $C_e$ is the channel size for each DB.

\begin{table}[!t]
    \caption{Temporally extended SNeRV architecture details.}
    \centering
    \vspace{-4mm}
    \begin{subtable}{0.45\textwidth}
        \centering
        \caption{Temporally extended SNeRV decoder architecture details.}
        \resizebox{4.5cm}{!}{%
        {\
        \begin{tabular}{l||ccc}
        \toprule
        Input size & Strides & $C_{0}$ & $N_{RB}$ \\ \midrule
        640$\times$1280   & 5, 4, 2, 2, 2        & 103 & 6                 \\
        480$\times$960    & 5, 3, 2, 2, 2        & 113 & 6                 \\
        960$\times$1920   & 5, 4, 3, 2, 2        & 96 & 6                 \\ \bottomrule
        \end{tabular}%
        }}
        \label{tab:SNeRV_Tdec_details}
    \end{subtable}
    \hfill
    \begin{subtable}{0.5\textwidth}
        \centering
        \caption{Temporally extended SNeRV encoder architecture details.}
        \resizebox{5cm}{!}{%
        \begin{tabular}{l||ccc}
        \toprule
        Input Size                & $e_{\Delta t}$ & Strides & $C_e$   \\ \midrule
        \multirow{3}{*}{640$\times$1280} & 3$\times$5$\times$10     & 4, 2, 2, 2, 2  & 64 \\
        & 3$\times$10$\times$20     & 4, 2, 2, 2  & 64 \\
                                  & 3$\times$20$\times$40     & 2, 2, 2, 2  & 64 \\ \midrule
        \multirow{3}{*}{480$\times$960}  & 3$\times$5$\times$10     & 2, 2, 2, 2, 2  & 64 \\
        & 3$\times$10$\times$20     & 2, 2, 2, 2  & 64 \\
                                  & 3$\times$20$\times$40     & 2, 2, 2     & 64 \\ \midrule
        \multirow{3}{*}{960$\times$1920} & 3$\times$5$\times$10     & 4, 3, 2, 2, 2  & 64 \\
        & 3$\times$10$\times$20     & 4, 3, 2, 2  & 64 \\
                                  & 3$\times$20$\times$40     & 3, 2, 2, 2  & 64 \\ \bottomrule
        \end{tabular}%
        }
        \label{tab:SNeRV_Tenc_details}
    \end{subtable}
    \label{tab:SNeRV_T_details}
    \vspace{-2mm}
\end{table}

\section{Ablation Study}
\subsection{Additional Ablation Results}
Detailed results of the ablation studies on the Bunny dataset are shown in \cref{tab:supp_ablation}, limited to a size of 3M. Also, in \cref{tab:supp_ablation_loss}, we present the results of different loss terms, when changing $\alpha$ in \cref{eq:loss}.

\subsection{Ablation Results for Different Model Sizes}
We present the regression performance, when increasing the model capacity. The quality of the reconstructed frames improves with an increase in the model capacity size, as demonstrated in \cref{tab:supp_ablation_size}.

\begin{table}[!ht]
    \centering
    \caption{Results of ablation studies.}
    \vspace{-4mm}
    \begin{subtable}{0.2\textwidth}
        \centering
        \caption{Ablation for number of residual blocks of MFBs in Ours(B).}
        \vspace{-1mm}
        \resizebox{2.5cm}{!}{%
        \begin{tabular}{c||cc}
        \toprule
        $N_{RB}$ & PSNR  & MS-SSIM   \\ \midrule
        4                & 39.48 & 0.9929 \\
        6                & 39.64 & 0.9928 \\
        8                & 39.70 & 0.9930 \\ \bottomrule
        \end{tabular}%
        }
        \label{tab:ablation_rbnum}
    \end{subtable}
    \hfill
    \begin{subtable}{0.3\textwidth}
        \centering
        \caption{Ablation for expansion rate of TUBs in Ours(T).}
        \vspace{2mm}
        \resizebox{4cm}{!}{%
        \begin{tabular}{c||cc}
        \toprule
        Expansion of TUB & PSNR  & MS-SIM   \\ \midrule
        $\times1$               & 38.88 & 0.9909 \\
        $\times2$               & 39.09 & 0.9913 \\
        $\times3$               & 38.95 & 0.9910 \\ \bottomrule
        \end{tabular}%
        }
        \label{tab:ablation_tubexp}
    \end{subtable}
    \hfill
    \begin{subtable}{0.3\textwidth}
        \centering
        \caption{Ablation study on different loss terms.}
        \resizebox{4.5cm}{!}{%
        \begin{tabular}{l||cc|cc}
        \toprule
        \multirow{2}{*}{Loss term} & \multicolumn{2}{c|}{Ours(B)} & \multicolumn{2}{c}{Ours(T)} \\ \cmidrule{2-5}
         & PSNR         & MS-SSIM          & PSNR         & MS-SSIM         \\ \midrule
        L2            & 38.35        & 0.9878        & 37.95        & 0.9860        \\
        L1 ($\alpha=1$)             & 39.66        & 0.9920         & 38.99        & 0.9898       \\
        $\alpha=0.7$       & 39.64        & 0.9928        & 39.09        & 0.9913       \\
        $\alpha=0.5$       & 39.36        & 0.9927        & 38.73        & 0.9912       \\ \bottomrule
        \end{tabular}%
        }
        \label{tab:supp_ablation_loss}
    \end{subtable}
    \label{tab:supp_ablation}
    \vspace{-2mm}
\end{table}

\begin{table}[!ht]
\centering
    \caption{Detailed model architecture for different model sizes. $e_{\Delta t}$ is not used in Ours(B).}
    \resizebox{0.7\columnwidth}{!}{%
    \begin{tabular}{c||ccccc|cc}
    \toprule
    Model                    & Total size & Decoder size & $C_0$ & $N_{RB}$ & $e_{\Delta t}$ & PSNR  & MS-SSIM   \\ \midrule
    \multirow{6}{*}{Ours(B)} & 0.34M          & 0.36M            & 41     & 2 & -   & 30.88 & 0.9555 \\
                             & 0.75M          & 0.76M            & 57     & 3  & -       & 33.25 & 0.9742 \\
                             & 1.50M          & 1.51M            & 81     & 4  & -        & 36.76 & 0.9871 \\
                             & 2.99M          & 3.01M            & 111     & 6 & -         & 39.64 & 0.9928 \\
                             & 4.57M          & 4.55M            & 132     & 8  & -        & 41.24 & 0.9949 \\
                             & 5.95M          & 5.93M            & 148     & 10 & -        & 41.66 & 0.9953 \\ \midrule
    \multirow{6}{*}{Ours(T)} & 0.30M         & 0.36M            & 39     & 2    & 3$\times$5$\times$10  & 30.40 & 0.9479 \\
                             & 0.69M          & 0.75M            & 61     & 3  & 3$\times$5$\times$10   & 33.19 & 0.9722 \\
                             & 1.39M          & 1.57M            & 82     & 4  & 3$\times$10$\times$20   & 36.53 & 0.9860 \\
                             & 2.34M         & 2.99M            & 103     & 6 & 3$\times$20$\times$40         & 39.09 & 0.9913 \\
                             & 4.52M          & 3.87M            & 129     & 8 & 3$\times$20$\times$40         & 40.89 & 0.9942 \\
                             & 5.95M          & 5.30M            & 146     & 10 & 3$\times$20$\times$40       & 42.02 & 0.9954 \\ \bottomrule
    \end{tabular}%
    }
    \label{tab:supp_ablation_size}
\end{table}
\vspace{-2mm}
\begin{table}[!th]
\caption{Detailed model architecture for video compression.}
\centering
    \resizebox{0.5\columnwidth}{!}{%
    \begin{tabular}{l||cc|cc}
    \toprule
    Model                    & Total size  & Decoder size & $C_{0}$  & $N_{RB}$ \\ \midrule
    \multirow{3}{*}{Ours(B)}  & 2.75M       & 2.67M         & 95  & 6         \\
                             & 4.32M       & 4.24M         & 120 & 6         \\
                             & 6.64M       & 6.71M         & 150 & 6         \\ \bottomrule
    \end{tabular}%
    }
    \label{tab:compression_details}
\end{table}

\begin{figure}[!tb]
\centering
   \includegraphics[width=0.9\textwidth]{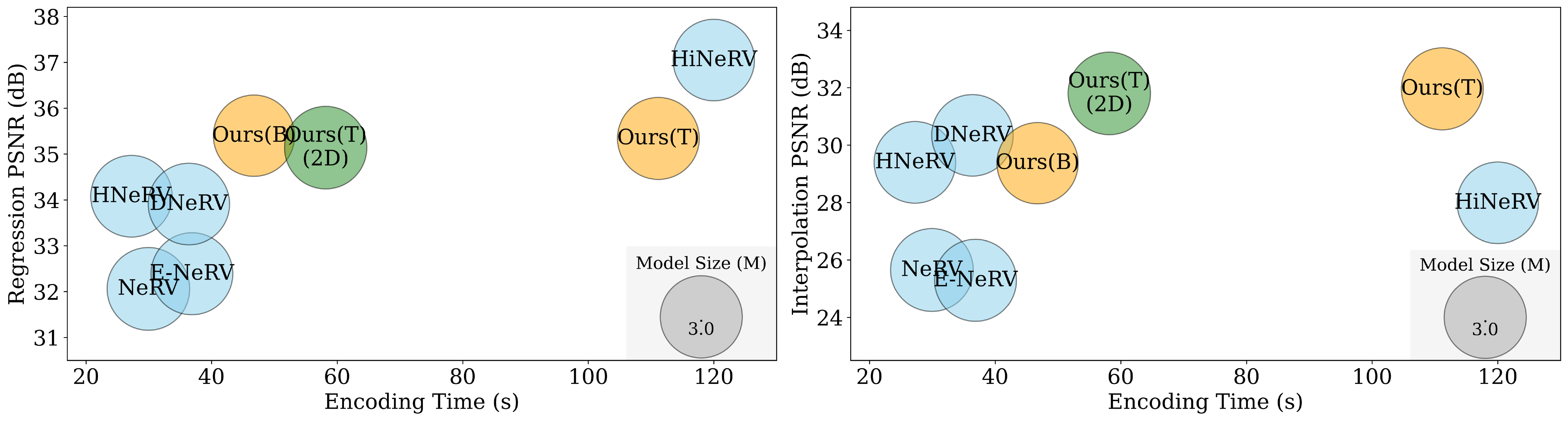}
\caption{Encoding complexity comparisons in UVG datasets.}
\label{fig:supp_enctime}
\end{figure}

\section{Complexity}
\subsection{Encoding Complexity}
In \cref{fig:supp_enctime}, we present the total encoding (training) time per frame in regression and interpolation along with the performance. We used 3M-sized models and NVIDIA RTX 3090 GPU in the tests. Ours(B) consumes 46.7(s), while HNeRV\cite{Hnerv} and DNeRV\cite{zhao2023dnerv} require 27.2(s) and 36.4(s) to encode the UVG datasets, respectively. In regression, Ours(B) offers a  reasonable trade-off compared to the state-of-the-art studies. The encoding time is comparable with the MACs reported in the paper, while some customized optimization can be applied.

\subsection{Encoding Complexity}
In Ours(T), the embedding for each time step is processed with Conv3D, which is complex. We reduce the complexity through replacing Conv3D with Conv2D by concatenating the time axis into channel axis and display the results in \cref{fig:supp_enctime} as Ours(T)(2D). The encoding time and MACs of Ours(T) are reduced from 111.2(s) and 401.57G to 58.2(s) and 254.47G, respectively, with a slight drop in performance.

\section{Video Compression}
\subsection{Experimental Details for SNeRV Compression} \label{sec:compression}
For video compression, we use three different sizes of models, each corresponding to a single bpp. We report the details of models used in the experiment in \cref{tab:compression_details}. The first UB's input channel size, $C_0$, and number of residual blocks used in the MFBs are adjusted to achieve different model sizes. If not specified, we use 10\% model pruning and 8-bit quantization for decoder parameters and 6-bit quantization for embeddings to compress the model. 

\subsection{Additional Results for Compression}
Detailed coding performance for our backbone is reported in \cref{tab:compression_ours}. In \cref{tab:supp_nncodec_uvg}, we show the Bjontegaard Delta (BD) rate in terms of PSNR, using our compressed backbone as an anchor, which is explained in \cref{sec:compression}. As aforementioned in the main manuscript, we used the same coding pipeline (i.e., 10$\%$ pruning and quantization) of the previous studies for fair comparisons. \cref{tab:supp_nncodec_uvg} presents the coding performance of the model as compared to non-compressed model denoted by ``No compression.'' As shown in the table, the model compression reduces its original size from approximately one half to one third. 

The superior performance of SNeRV is largely due to the effectiveness of the proposed modules, successfully reconstructing detailed high-frequency components. This approach is orthogonal to other advanced coding schemes in quantization and entropy coding.
We further test the coding performance, when directly employing NNCodec\cite{becking2023nncodec} as an international standard on compression of neural networks to compress the decoders of the SNeRV. The results by NNCodec with the default quantization parameter (QP) value of -38 are presented in \cref{tab:supp_nncodec_uvg}. Utilizing NNcodec results in average -14.10$\%$ BD-rate saving on UVG datasets. The BD-rate saving of -37.70$\%$ is observed in the Bee sequence. This result indicates the potential coding performance of our model. When directly applying a few more coding schemes, such as adaptive arithmetic entropy coding, the coding efficiency of SNeRV was considerably improved.


The comparisons of VMAF\cite{rassool2017vmaf} scores for the compression results are demonstrated in \cref{tab:compression_vmaf}.

\begin{table}[!ht]
\centering
\caption{Compression results of SNeRV on UVG dataset.}
    \resizebox{0.7\columnwidth}{!}{%
    \begin{tabular}{l||c|ccccccc|c}
    \toprule
    Model & bpp & Beauty & Bosph & Bee & Jockey & Ready & Shake & Yacht & Avg. \\ \midrule
    \multirow{3}{*}{Ours(B)} & 0.051  & 34.19  & 36.91 & 38.83 & 35.42  & 28.89 & 35.90 & 31.09 & 34.46 \\
       & 0.033  & 34.02  & 35.74 & 38.45 & 34.09  & 27.17 & 35.14 & 29.93 & 33.51  \\
       & 0.021  & 33.78  & 33.97 & 38.19 & 32.51 & 25.49  & 34.46 & 28.74 & 32.45  \\ \bottomrule
    \end{tabular}%
    }
    \label{tab:compression_ours}
\end{table}

\begin{table}[!ht]
\centering
\caption{BD-rate results on UVG datasets, when using the compressed backbone presented in the main manuscript as an anchor.}
    \resizebox{\columnwidth}{!}{%
    \begin{tabular}{l||ccccccc|c}
    \toprule
    Method         & Beauty   & Bosph    & Bee      & Jockey   & Ready    & Shake    & Yacht    & Avg.     \\ \midrule
    No compression & 255.27\% & 269.95\% & 184.18\% & 281.84\% & 306.36\% & 258.37\% & 289.05\% & 263.57\% \\ \midrule
    Ours(B)+NNC    & -10.39\% & -11.84\% & -37.70\% & -8.53\%  & -4.30\%  & -18.85\% & -7.12\%  & -14.10\% \\ \bottomrule
    \end{tabular}%
    }
    \label{tab:supp_nncodec_uvg}
\end{table}

\begin{table}[!thb]
\centering
\caption{VMAF scores for compression results.}
    \resizebox{\textwidth}{!}{%
    \begin{tabular}{l||cccc|cccc|cccc}
    \toprule
    Method & HNeRV\cite{Hnerv} & DNeRV\cite{zhao2023dnerv} & H.265\cite{sullivan2012overview} & Ours(B) & HNeRV\cite{Hnerv} & DNeRV\cite{zhao2023dnerv} & H.265\cite{sullivan2012overview} & Ours(B) & HNeRV\cite{Hnerv} & DNeRV\cite{zhao2023dnerv} & H.265\cite{sullivan2012overview} & Ours(B) \\ \midrule
    bpp    & 0.023     & 0.023     & 0.019     & 0.021  & 0.037     & 0.034     & 0.038     & 0.033  & 0.053    & 0.050     & 0.059     & 0.051 \\ \midrule
    Beauty & 71.42 & 70.51 & 64.90 & 77.22 & 77.94 & 73.53 & 71.43 & 81.67 & 82.42 & 76.25 & 73.36 & 83.69 \\
    Bosph  & 60.75 & 54.96 & 69.46 & 72.43 & 70.81 & 59.41 & 77.12 & 78.73 & 76.98 & 67.68 & 80.83 & 83.76 \\
    Bee    & 93.78 & 93.04 & 85.27 & 93.28 & 94.65 & 93.27 & 86.57 & 93.40 & 94.69 & 93.88 & 87.58 & 94.07 \\
    Jockey & 54.38 & 61.73 & 84.22 & 65.25 & 64.49 & 64.78 & 93.79 & 75.53 & 73.10 & 77.72 & 96.26 & 85.37 \\
    Ready  & 39.52 & 47.52 & 72.16 & 55.43 & 48.20 & 49.30 & 87.47 & 64.83 & 55.97 & 62.89 & 93.48 & 73.82 \\
    Shake  & 69.49 & 68.33 & 49.30 & 70.05 & 75.07 & 70.06 & 59.43 & 74.75 & 79.53 & 70.55 & 65.21 & 80.08 \\
    Yacht  & 44.46 & 42.24 & 48.20 & 53.70 & 50.72 & 44.04 & 55.50 & 60.23 & 56.91 & 53.12 & 62.28 & 67.4 \\ \midrule
    Avg.   & 61.97 & 62.62 & 67.65 & 69.62 & 68.84 & 64.91 & 75.90 & 75.59 & 74.23 & 71.73 & 79.86 & 81.17 \\ \bottomrule
    \end{tabular}%
    }
    \label{tab:compression_vmaf}
\end{table}

\section{Quantitative Results}
Performance evaluation on video regression and interpolation for 20 subsets of DAVIS datasets are reported in \cref{tab:supp_davis}.
We also provide detailed results for video in-painting on DAVIS datasets in \cref{tab:supp_davis_inp}.

\begin{table}[!thb]
\centering
\caption{Quantitative comparisons of regression and interpolation tasks on DAVIS datasets (PSNR/MS-SSIM).}
    \resizebox{\textwidth}{!}{%
    \begin{tabular}{l||cccc|cccc}
    \toprule
    \multirow{2}{*}{Video} & \multicolumn{4}{c|}{Regression}                           & \multicolumn{4}{c}{Interpolation}                         \\
                           & HNeRV\cite{Hnerv}        & DNeRV\cite{zhao2023dnerv}        & Ours(B)      & Ours(T)      & HNeRV\cite{Hnerv}        & DNeRV\cite{zhao2023dnerv}        & Ours(B)      & Ours(T)      \\ \midrule
    bike-packing           & 30.75/0.9618 & 31.10/0.9635 & 33.29/0.9766 & 31.90/0.9677 & 19.26/0.6972 & 20.39/0.7624 & 19.48/0.7034 & 20.47/0.7461 \\
    blackswan              & 31.77/0.9545 & 32.61/0.9576 & 33.83/0.9708 & 33.08/0.9621 & 20.77/0.5391 & 21.71/0.6213 & 20.80/0.5361 & 22.19/0.6229 \\
    breakdance             & 28.84/0.9729 & 29.52/0.9755 & 31.40/0.9827 & 31.26/0.9809 & 20.33/0.8566 & 21.51/0.8707 & 20.23/0.858  & 21.87/0.8813 \\
    breakdance-flare & 28.84/0.8872 & 28.71/0.8787 & 30.36/0.9222 & 29.71/0.909  & 20.70/0.5396 & 20.96/0.5233 & 20.80/0.5529 & 22.55/0.6326 \\
    camel                  & 26.59/0.9006 & 26.97/0.9093 & 28.68/0.9331 & 30.14/0.948  & 19.02/0.5359 & 19.3/0.6434  & 19.10/0.5335 & 22.37/0.7672 \\
    car-roundabout   & 28.28/0.9445 & 29.26/0.9551 & 31.19/0.9677 & 30.72/0.9646 & 16.17/0.5717 & 19.57/0.7403 & 16.27/0.5773 & 19.17/0.7238 \\
    car-shadow             & 33.50/0.9633 & 34.36/0.9712 & 35.79/0.9742 & 34.33/0.9677 & 18.51/0.6307 & 21.25/0.7706 & 18.39/0.6317 & 21.31/0.7786 \\
    car-turn               & 29.88/0.9227 & 30.68/0.9273 & 32.12/0.9463 & 31.45/0.9308 & 22.01/0.6377 & 22.92/0.7122 & 21.99/0.6356 & 23.42/0.7053 \\
    cows                   & 23.85/0.8328 & 23.64/0.8314 & 25.14/0.8857 & 27.38/0.9383 & 18.87/0.5322 & 18.11/0.5116 & 18.82/0.5317 & 21.64/0.7235 \\
    dance-jump             & 31.27/0.9334 & 31.59/0.9371 & 33.03/0.9592 & 31.80/0.9372 & 20.90/0.5429  & 21.10/0.5836 & 20.77/0.5361 & 22.03/0.6151 \\
    dance-twirl            & 28.55/0.8877 & 28.89/0.8997 & 30.41/0.9310 & 29.93/0.9199 & 17.56/0.4797 & 18.96/0.5809 & 17.63/0.4701 & 19.12/0.5704 \\
    dog-agility            & 35.74/0.9869 & 35.86/0.9871 & 36.88/0.9900 & 35.19/0.9841 & 16.91/0.6415 & 19.25/0.6782 & 17.03/0.6519 & 18.49/0.6862 \\
    drift-chicane          & 41.42/0.9908 & 42.94/0.9930 & 42.96/0.9933 & 41.38/0.9925 & 34.31/0.9685 & 34.38/0.9692 & 33.97/0.9689 & 34.86/0.9727 \\
    elephant               & 28.65/0.9147 & 29.18/0.9251 & 30.84/0.9474 & 32.16/0.9565 & 22.52/0.7179 & 21.87/0.7102 & 22.40/0.7089 & 25.48/0.8304 \\
    flamingo               & 29.37/0.9167 & 30.53/0.9297 & 32.15/0.9497 & 32.51/0.9494 & 20.89/0.6420 & 21.10/0.6742 & 20.63/0.6308 & 23.19/0.7328 \\
    goat                   & 26.84/0.9094 & 27.25/0.9100 & 29.13/0.9461 & 26.23/0.8863 & 16.61/0.2610 & 17.97/0.4404 & 16.69/0.2561 & 17.44/0.3435 \\
    mallard-water          & 30.23/0.9475 & 30.44/0.9506 & 32.46/0.9711 & 29.99/0.9486 & 16.63/0.3760 & 18.49/0.5773 & 16.87/0.3749 & 18.51/0.5171 \\
    parkour                & 25.62/0.8668 & 27.08/0.8986 & 28.77/0.9290 & 27.16/0.8954 & 17.58/0.4347 & 20.39/0.6452 & 17.93/0.4727 & 20.80/0.6398 \\
    scooter-black          & 29.58/0.9665 & 29.74/0.9678 & 31.71/0.9777 & 28.83/0.9620 & 13.58/0.4147 & 16.23/0.6244 & 13.62/0.4134 & 14.71/0.5219 \\
    stroller               & 32.68/0.9474 & 33.41/0.9555 & 35.66/0.9750 & 32.23/0.9469 & 19.63/0.5129 & 22.02/0.6771 & 19.83/0.5161 & 21.14/0.5893 \\ \midrule
    Avg.                   & 30.11/0.9304 & 30.69/0.9362 & 32.29/0.9564 & 31.37/0.9474 & 19.64/0.5766 & 20.87/0.6658 & 19.66/0.5780 & 21.54/0.6800 \\ \bottomrule
    \end{tabular}%
    }
    \label{tab:supp_davis}
\end{table}

\begin{table}[!thb]
\centering
\caption{Quantitative comparisons of in-painting task on DAVIS datasets (PSNR/MS-SSIM).}
    \resizebox{0.9\textwidth}{!}{%
    \begin{tabular}{l||ccc|ccc}
    \toprule
    \multirow{2}{*}{Video} & \multicolumn{3}{c|}{In-painting1}         & \multicolumn{3}{c}{In-painting2}          \\
                           & HNeRV        & DNeRV        & Ours(B)      & HNeRV        & DNeRV        & Ours(B)      \\ \midrule
    bike-packing           & 30.32/0.9594 & 30.25/0.9599 & 31.16/0.9713 & 30.09/0.9575 & 29.94/0.9564 & 31.00/0.9687 \\
    blackswan              & 31.52/0.9516 & 32.04/0.9548 & 33.19/0.9674 & 31.18/0.9503 & 31.97/0.9542 & 32.13/0.9647 \\
    breakdance             & 28.26/0.9693 & 28.51/0.9713 & 28.44/0.9752 & 27.51/0.9660 & 28.02/0.9680 & 27.63/0.9715 \\
    breakdance-flare       & 28.61/0.8811 & 28.28/0.8757 & 30.11/0.9216 & 28.45/0.8825 & 28.41/0.8761 & 29.22/0.9175 \\
    camel                  & 26.46/0.8988 & 26.64/0.9071 & 28.21/0.9322 & 26.33/0.8986 & 26.68/0.9049 & 28.08/0.9303 \\
    car-roundabout         & 27.98/0.9421 & 28.86/0.9537 & 30.12/0.9653 & 27.66/0.9405 & 28.66/0.9517 & 29.62/0.9636 \\
    car-shadow             & 33.28/0.9649 & 33.19/0.9692 & 33.62/0.9719 & 32.23/0.9616 & 32.47/0.9673 & 30.63/0.9667 \\
    car-turn               & 29.46/0.9196 & 30.13/0.9246 & 30.63/0.9407 & 29.35/0.9107 & 29.98/0.9217 & 29.69/0.9368 \\
    cows                   & 23.73/0.8295 & 23.34/0.8279 & 24.98/0.8862 & 23.75/0.8297 & 23.55/0.8266 & 24.57/0.8790 \\
    dance-jump             & 31.05/0.9312 & 31.16/0.9352 & 32.54/0.9572 & 30.48/0.9296 & 30.83/0.9347 & 31.61/0.9558 \\
    dance-twirl            & 28.24/0.8864 & 28.35/0.8967 & 29.93/0.9292 & 27.83/0.8839 & 28.10/0.8952 & 29.45/0.9297 \\
    dog-agility            & 34.13/0.9842 & 34.32/0.9845 & 32.12/0.9837 & 33.88/0.9823 & 34.51/0.9832 & 32.13/0.9819 \\
    drift-chicane          & 40.66/0.9895 & 41.61/0.9913 & 32.78/0.9815 & 40.10/0.9890 & 41.07/0.9906 & 31.24/0.9741 \\
    elephant               & 28.54/0.9115 & 28.89/0.9232 & 30.38/0.9454 & 28.40/0.9104 & 28.90/0.9214 & 29.41/0.9418 \\
    flamingo               & 29.21/0.9142 & 29.95/0.9263 & 31.72/0.9505 & 29.30/0.9151 & 30.10/0.9255 & 31.49/0.9481 \\
    goat                   & 26.61/0.9055 & 26.78/0.9050 & 28.31/0.9376 & 26.43/0.9035 & 26.84/0.9030 & 28.05/0.9344 \\
    mallard-water          & 29.76/0.9435 & 29.66/0.9466 & 31.59/0.9682 & 29.15/0.9398 & 29.22/0.9441 & 30.60/0.9658 \\
    parkour                & 26.12/0.8813 & 26.72/0.8962 & 28.67/0.9321 & 26.10/0.8811 & 26.77/0.8938 & 28.19/0.9255 \\
    scooter-black          & 28.94/0.9645 & 28.81/0.9652 & 30.47/0.9765 & 27.95/0.9611 & 27.79/0.9606 & 28.80/0.9714 \\
    stroller               & 32.26/0.9453 & 32.73/0.9529 & 34.96/0.9729 & 32.04/0.9435 & 32.63/0.9521 & 34.54/0.9717 \\ \midrule
    Avg.                   & 29.76/0.9287 & 30.01/0.9334 & 30.70/0.9533 & 29.41/0.9272 & 29.82/0.9316 & 29.90/0.9500 \\ \bottomrule
    \end{tabular}%
    }
    \label{tab:supp_davis_inp}
\end{table}

\section{Qualitative Results}
\begin{figure}[!htb]
\centering
   \includegraphics[width=0.9\textwidth]{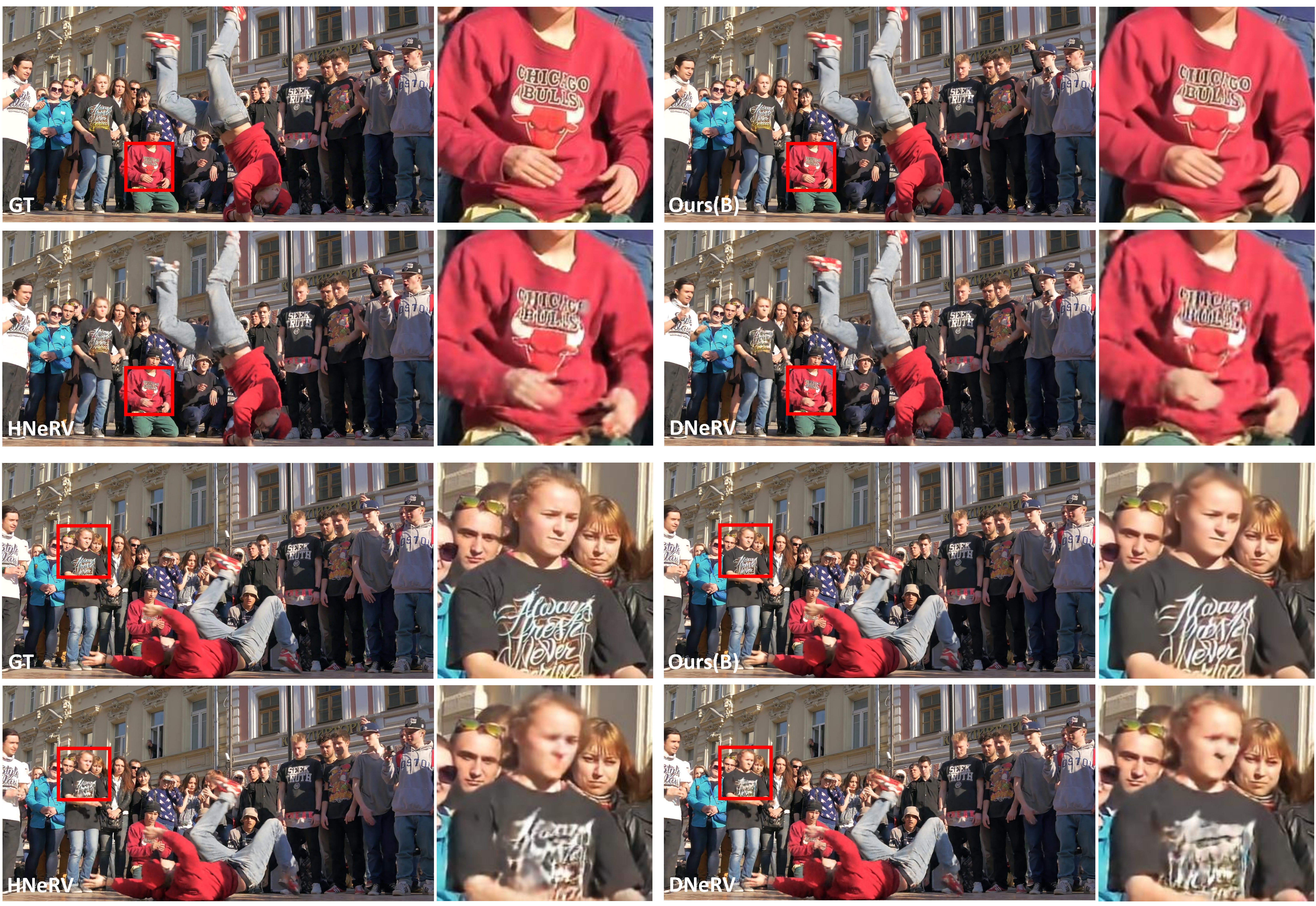}
\vspace{1mm}
\caption{Qualitative results of video regression task on Breakdance dataset at t=28 (top) and t=49 (bottom).}
\label{fig:supp_vis_reg1}
\vspace{-3mm}
\end{figure}

\begin{figure}[!htb]
\centering
   \includegraphics[width=0.9\textwidth]{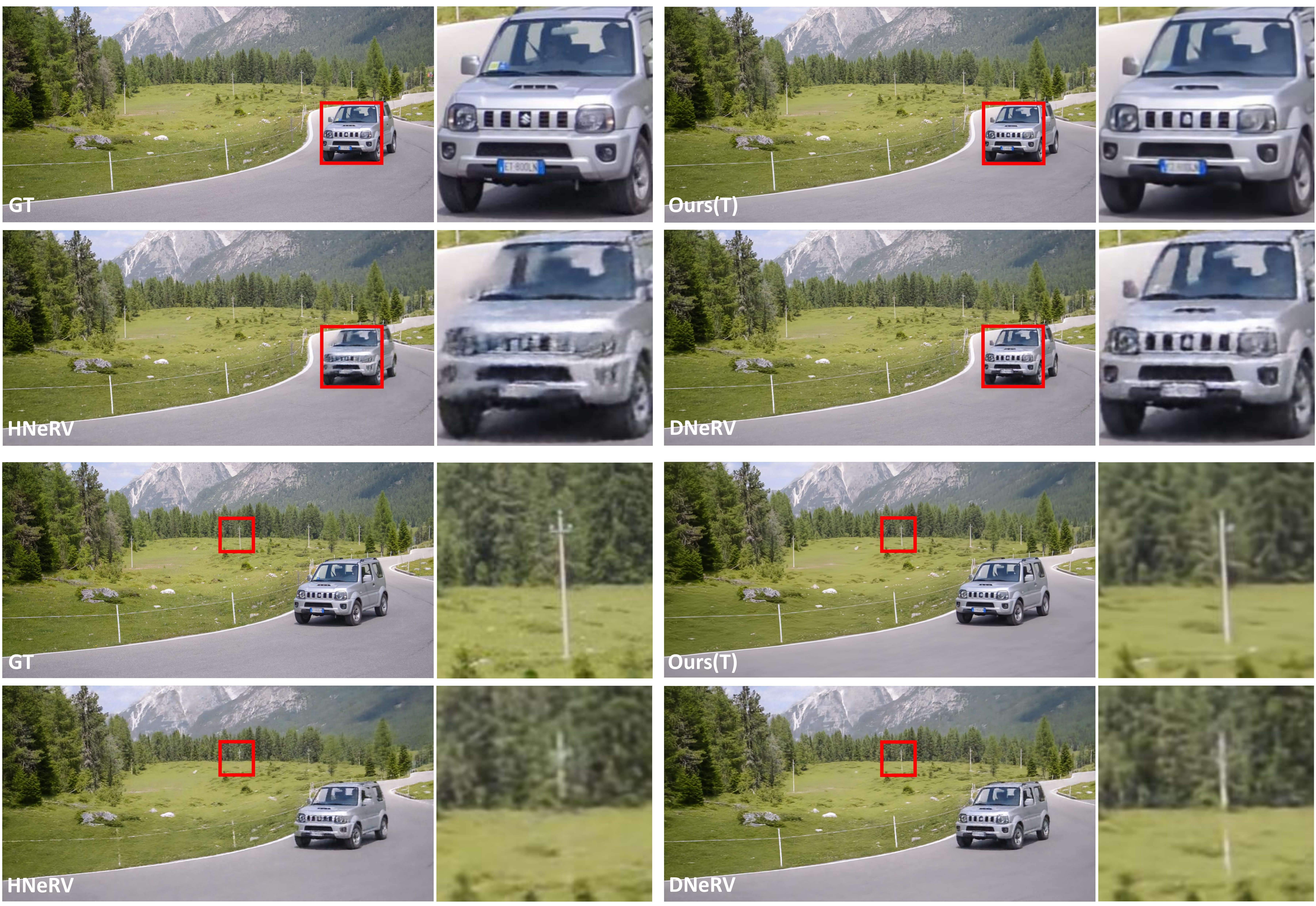}
\caption{Qualitative results of video regression task on Car-turn dataset at t=15 (top) and t=24 (bottom).}
\label{fig:supp_vis_reg2}
\vspace{-3mm}
\end{figure}

\begin{figure}[!htb]
\centering
   \includegraphics[width=0.9\textwidth]{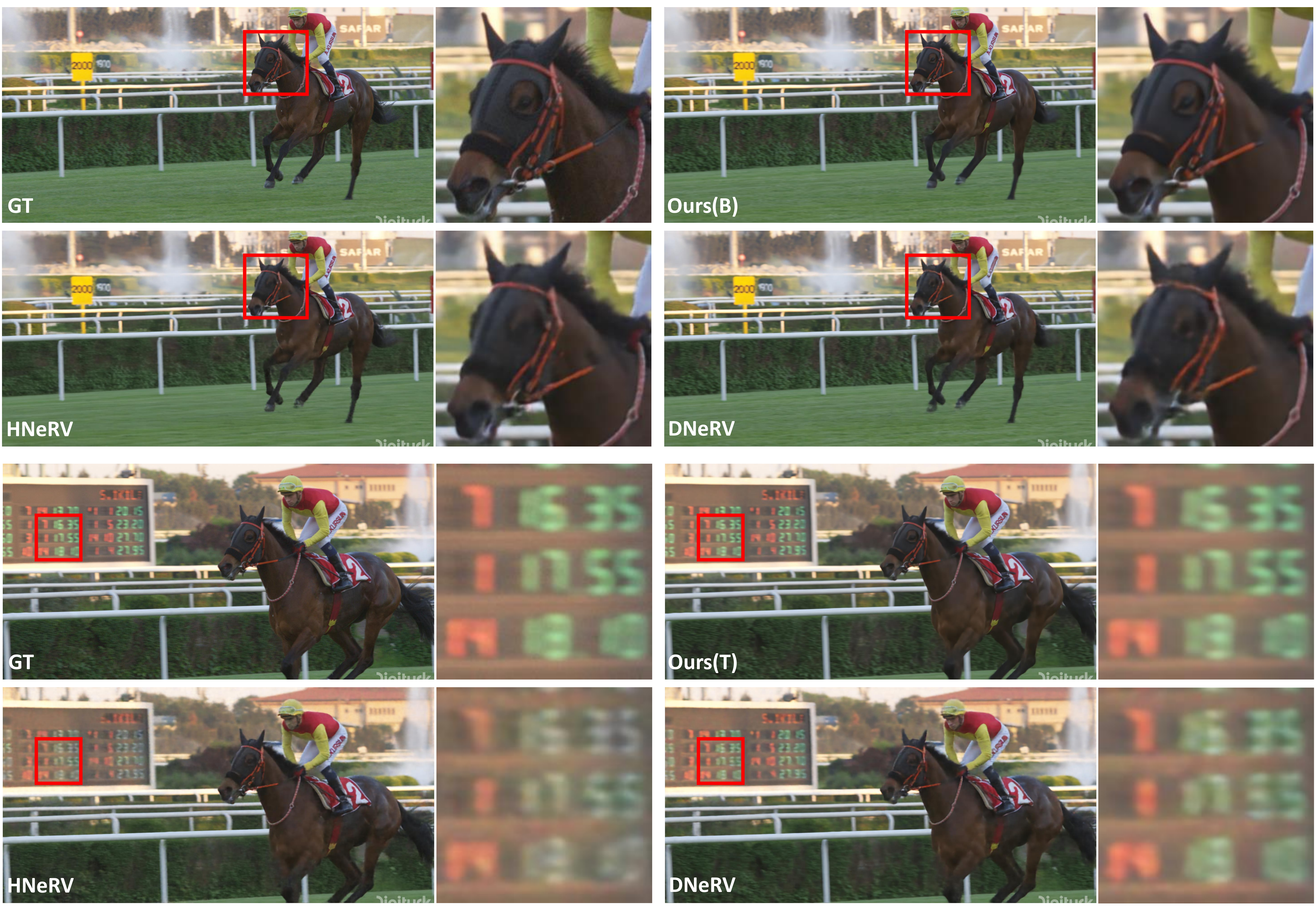}
\vspace{1mm}
\caption{Qualitative results of video regression task on Jockey dataset at t=85 (top) and t=179 (bottom).}
\label{fig:supp_vis_reg3}
\vspace{-3mm}
\end{figure}

\begin{figure}[!htb]
\centering
   \includegraphics[width=0.9\textwidth]{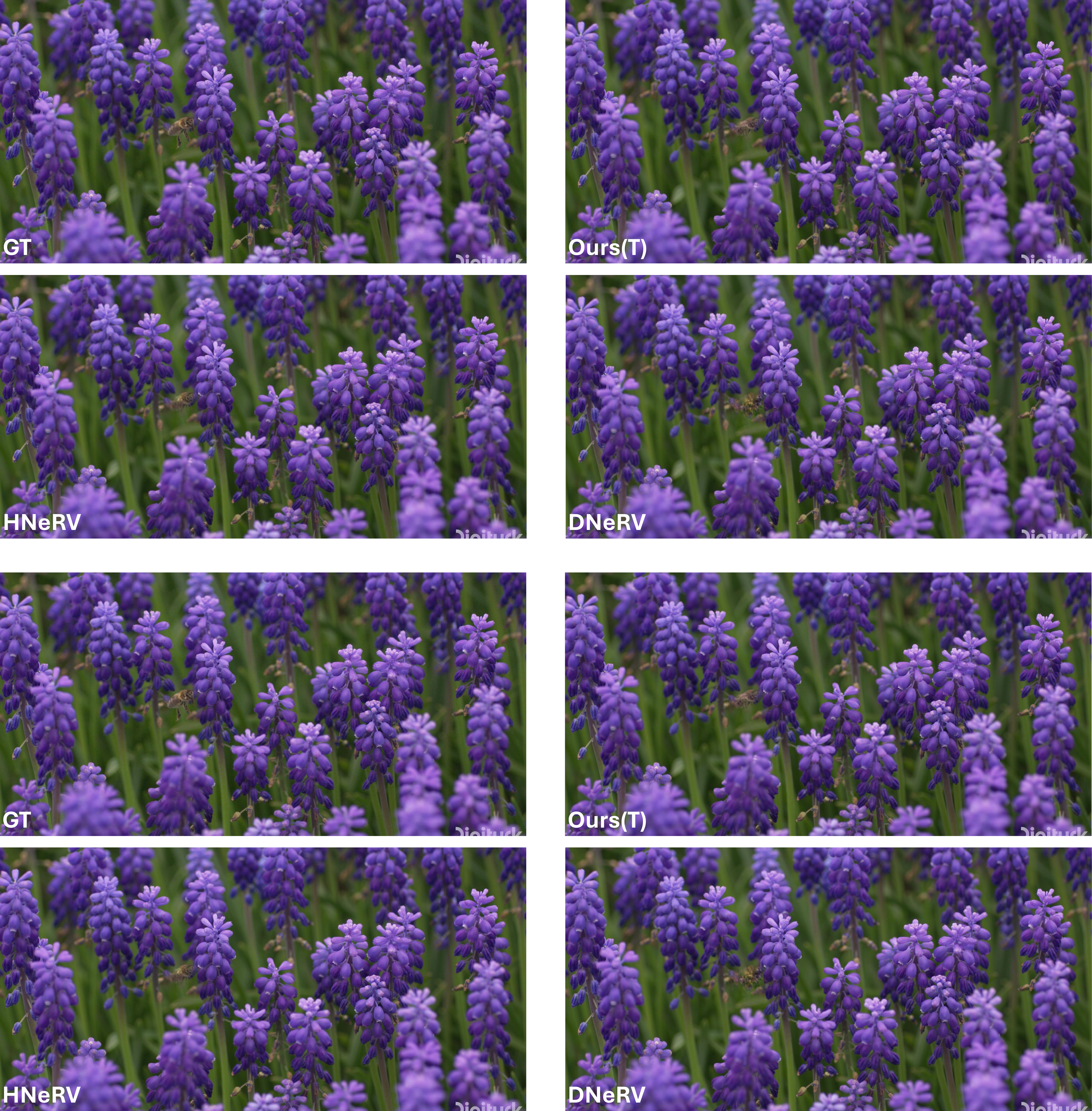}
\vspace{1mm}
\caption{Qualitative results of video interpolation task on Bee dataset at t=17 (top) and t=19 (bottom).}
\label{fig:supp_vis_interp1}
\vspace{-3mm}
\end{figure}

\begin{figure}[!htb]
\centering
   \includegraphics[width=0.9\textwidth]{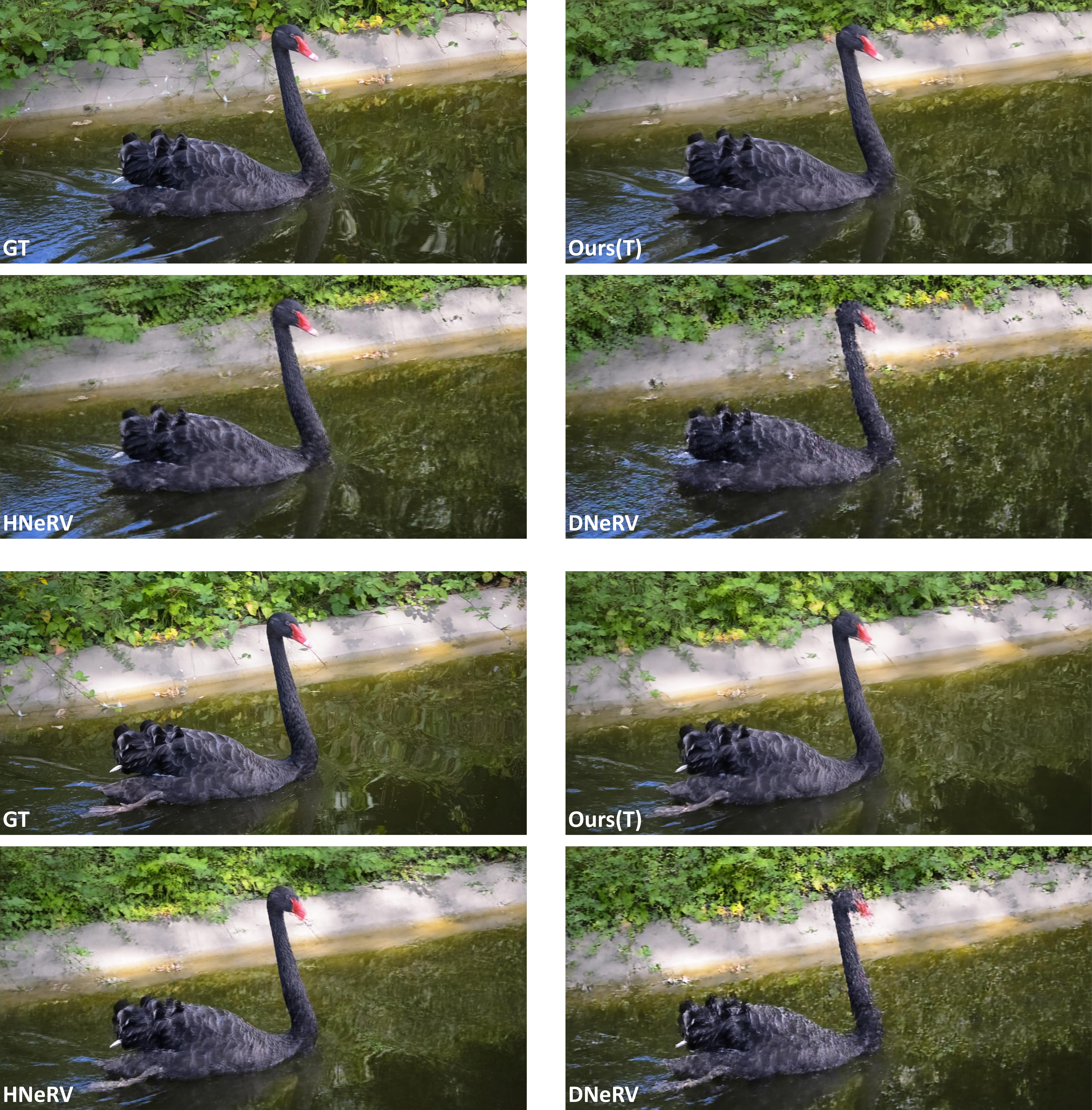}
\vspace{1mm}
\caption{Qualitative results of video interpolation task on Blackswan dataset at t=9 (top) and t=45 (bottom).}
\label{fig:supp_vis_interp2}
\vspace{-5mm}
\end{figure}

\begin{figure}[!htb]
\centering
   \includegraphics[width=0.9\textwidth]{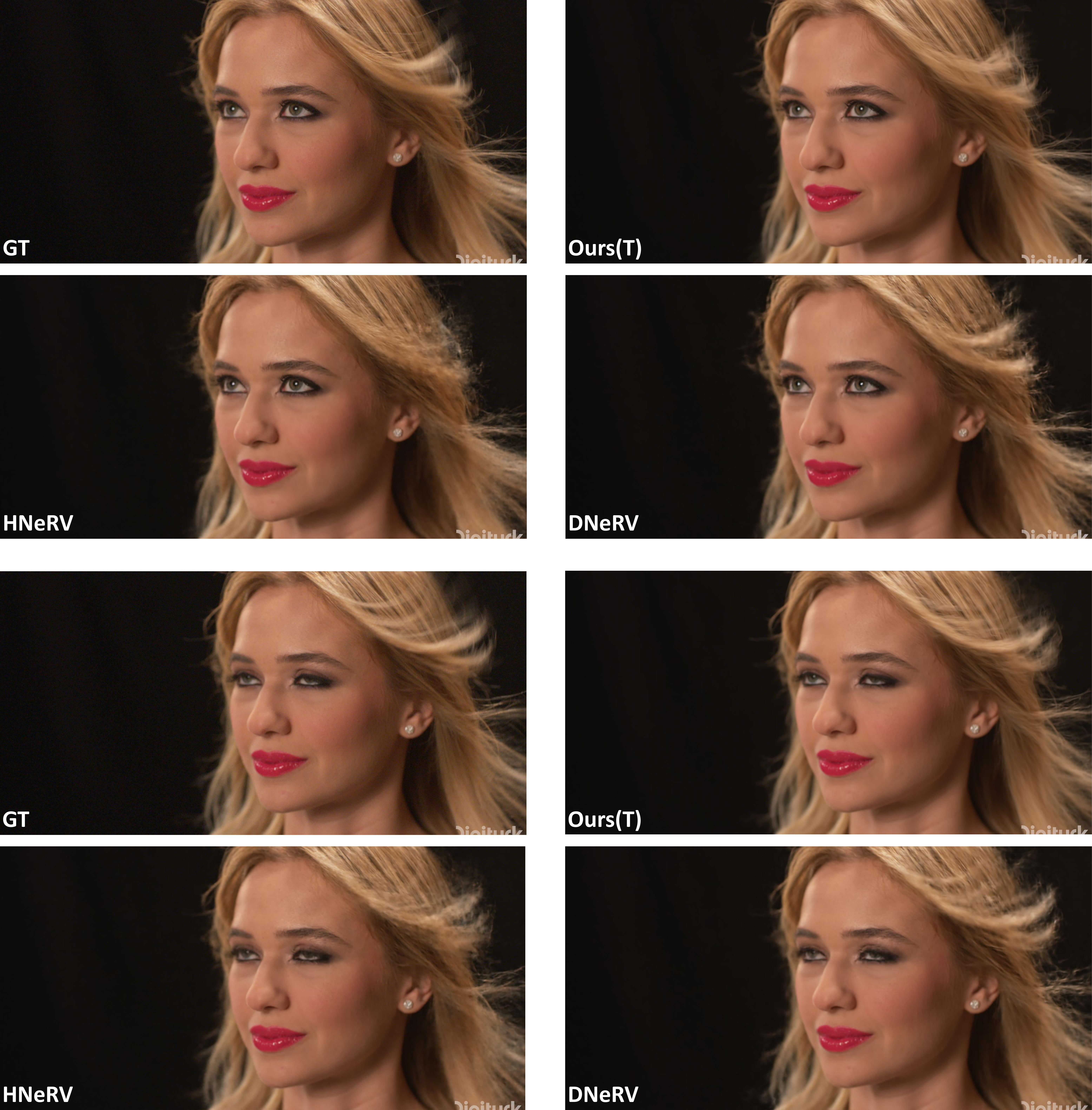}
\vspace{1mm}
\caption{Qualitative results of video interpolation task on Beauty dataset at t=83 (top) and t=171 (bottom).}
\label{fig:supp_vis_interp3}
\vspace{-5mm}
\end{figure}

\begin{figure}[!htb]
\centering
\begin{subfigure}{0.8\textwidth}
    \includegraphics[width=\columnwidth]{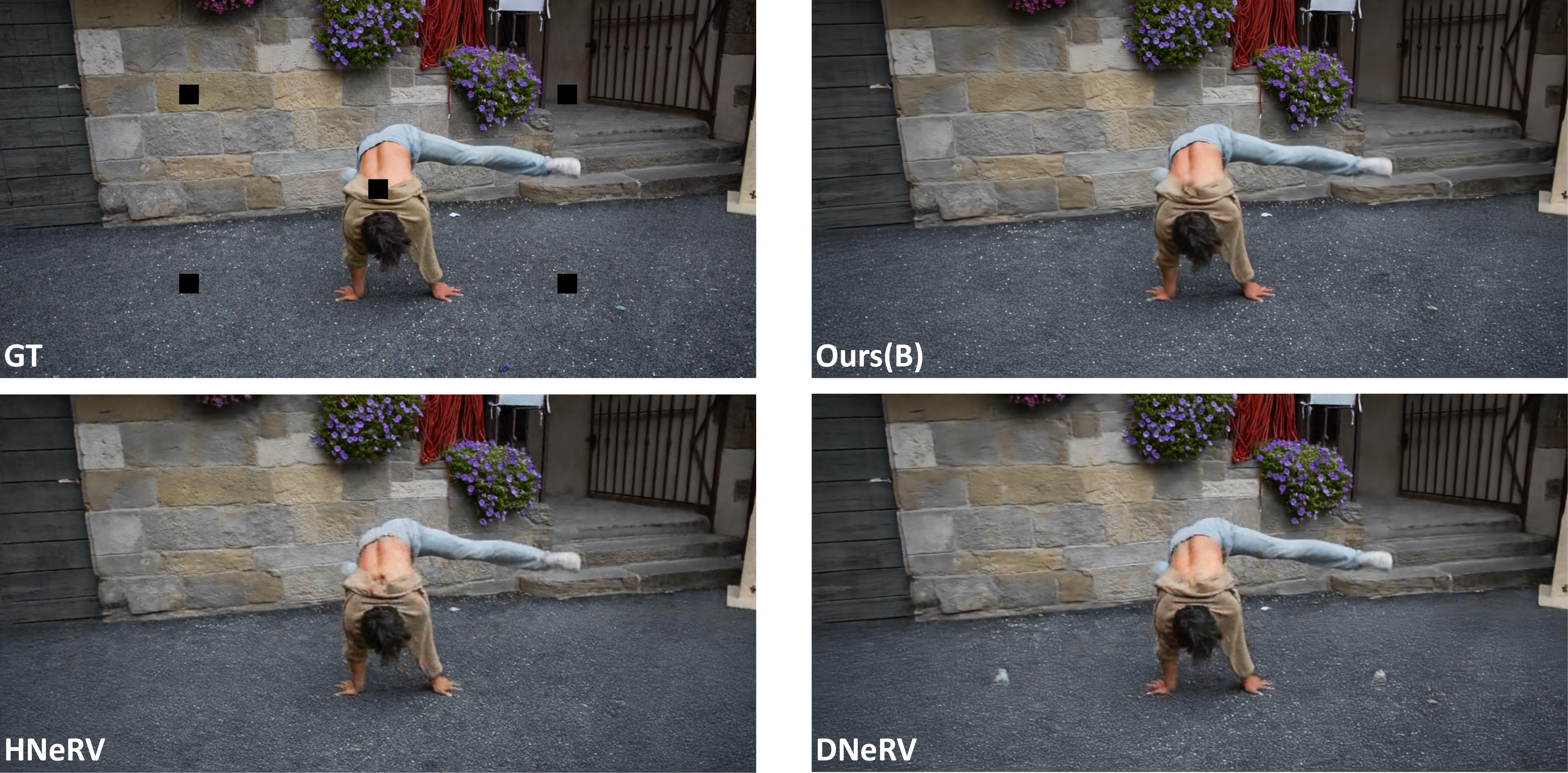}
    \caption{Qualitative results of video in-painting1 task on Breakdance-flare dataset}
    \vspace{2mm}
\end{subfigure}
\hfill
\begin{subfigure}{0.8\textwidth}
    \includegraphics[width=\columnwidth]{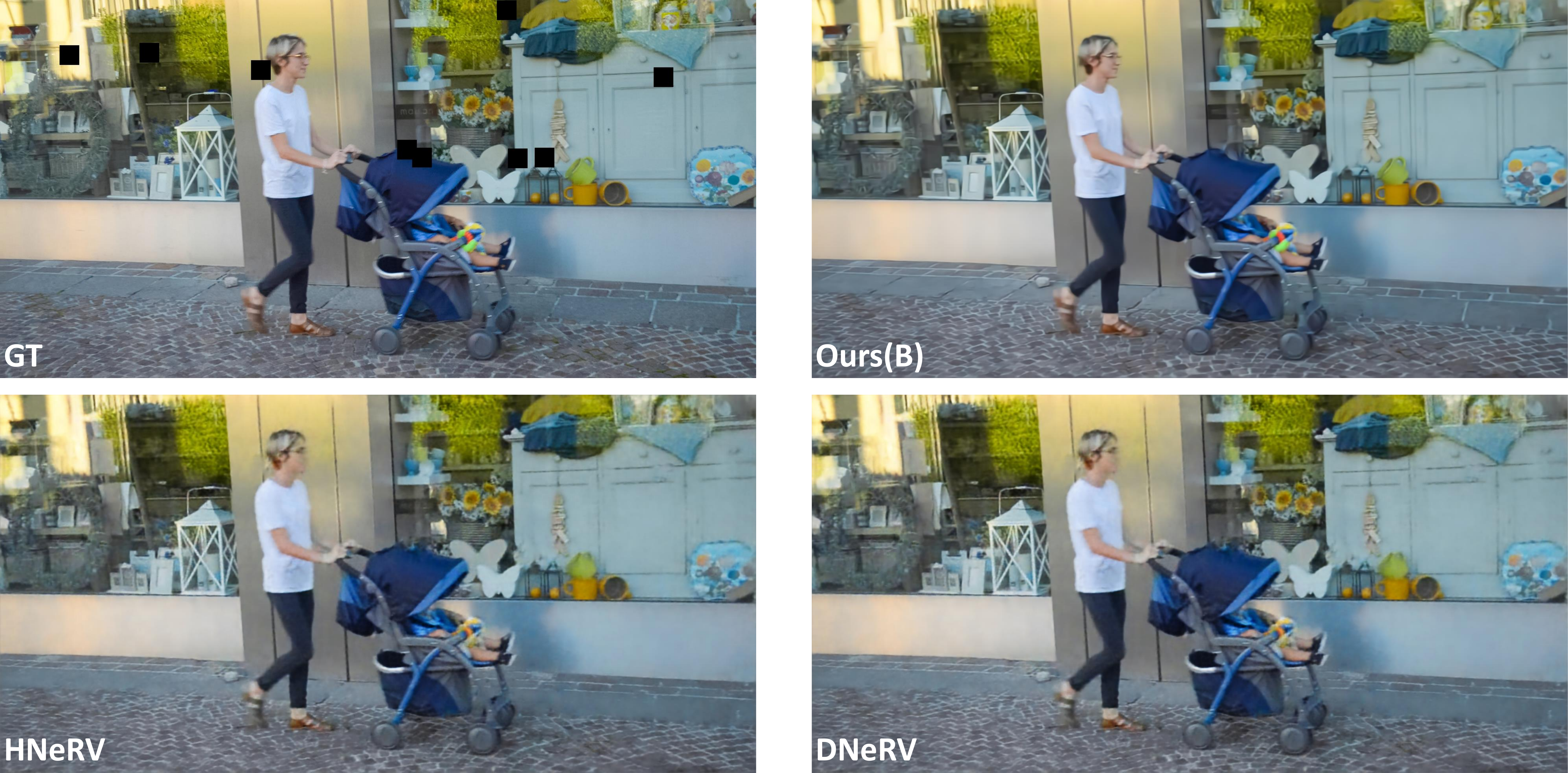}
    \caption{Qualitative results of video in-painting2 task on Stroller dataset}
    \vspace{2mm}
\end{subfigure}
\caption{Qualitative results of video in-painting tasks on DAVIS datasets.}
\label{fig:supp_vis_inp}
\vspace{-5mm}
\end{figure}

We provide additional qualitative comparisons on UVG and DAVIS datasets. The results for video regression can be found in \cref{fig:supp_vis_reg1}, \cref{fig:supp_vis_reg2}, and \cref{fig:supp_vis_reg3}. The results for video interpolation are provided in \cref{fig:supp_vis_interp1}, \cref{fig:supp_vis_interp2}, and \cref{fig:supp_vis_interp3}. Visualized video in-painting results for two different masks are in \cref{fig:supp_vis_inp}.

\end{document}